\documentclass[lettersize,journal]{IEEEtran}
\usepackage{amsmath,amsfonts,amssymb,amsthm}
\usepackage{algorithmic}
\usepackage{algorithm}
\usepackage{array}
\usepackage[caption=false,font=normalsize,labelfont=sf,textfont=sf]{subfig}
\usepackage{textcomp}
\usepackage{stfloats}
\usepackage{url}
\usepackage{verbatim}
\usepackage{graphicx}
\usepackage{cite}
\allowdisplaybreaks 

\usepackage{changes}

\def\BibTeX{{\rm B\kern-.05em{\sc i\kern-.025em b}\kern-.08em
    T\kern-.1667em\lower.7ex\hbox{E}\kern-.125emX}}

\usepackage{tikz}

\usetikzlibrary{arrows.meta, bending}

\usepackage{tikzscale}
\usepackage{pgfplots}
\pgfplotsset{compat=1.18} 
\usepgfplotslibrary{fillbetween}

\newtheorem*{thm@manual}{Theorem \manualthmnumber}
\newenvironment{manualthm}[1]{%
  \def\manualthmnumber{\ref{#1}}%
  \begin{thm@manual}%
}{%
  \end{thm@manual}%
}

\definecolor{tabblue}{HTML}{1f77b4}
\definecolor{taborange}{HTML}{ff7f0e}
\definecolor{tabgreen}{HTML}{2ca02c}
\definecolor{tabred}{HTML}{d62728}
\definecolor{tabpurple}{HTML}{9467bd}
\definecolor{tabbrown}{HTML}{8c564b}
\definecolor{tabpink}{HTML}{e377c2}
\definecolor{tabgray}{HTML}{7f7f7f}
\definecolor{tabolive}{HTML}{bcbd22}
\definecolor{tabcyan}{HTML}{17becf}
\definecolor{deepblue}{HTML}{0000ff}

\usepackage{mathtools}
\usepackage{bbm}

\usepackage[OT2,T1]{fontenc}

\newtheorem{theorem}{Theorem} 
\newtheorem{remark}{Remark} 
\newtheorem{lemma}{Lemma}
\newtheorem{proposition}{Proposition}
\newtheorem{corollary}{Corollary} 

\theoremstyle{definition}
\newtheorem{definition}{Definition}

\DeclarePairedDelimiterX{\infdivx}[2]{(}{)}{%
  #1\;\delimsize\|\;#2%
}
\DeclarePairedDelimiterX{\infdivmi}[2]{(}{)}{%
  #1;#2%
}
\DeclarePairedDelimiterX{\infdivcsmi}[2]{(}{)}{%
  #1 \to #2%
}
\DeclarePairedDelimiterX{\innerProd}[2]{\langle}{\rangle}{%
    #1,#2%
}
\DeclarePairedDelimiter{\abs}{\lvert}{\rvert}

\DeclarePairedDelimiterX{\SMI}[3]{I_{#1}^\text{\scriptsize s}(}{)}{#2; #3}
\DeclarePairedDelimiterX{\RFI}[3]{I_{F, #1}(}{)}{#2 \to #3}
\DeclarePairedDelimiterX{\RCSD}[3]{D_{#1}^{\mathrm{CS}}(}{)}{#2\;\delimsize\|\;#3}

\newcommand{\KLD}{D_{\mathrm{KL}}\infdivx}
\newcommand{\alphaD}[3]{D_{#1}\infdivx{#2}{#3}}

\newcommand{\MI}{I\infdivmi}
\newcommand{\CSD}{D_{\mathrm{CS}}\infdivx}

\newcommand{\qty}[1]{\left\{ #1 \right\}}
\newcommand{\iid}{\overset{\text{i.i.d.}}{\sim}}
\newcommand{\norm}[1]{\left\lVert #1 \right\rVert}
\newcommand{\length}[1]{\left\lvert #1 \right\rvert}
\newcommand{\RND}[2]{\frac{d #1}{d #2}}
\newcommand{\Ind}[1]{\mathbf{1}\!\left[ #1 \right]}
\newcommand{\Normal}[2]{\mathcal{N}(#1, #2)}
\newcommand{\FI}{I_F\infdivcsmi}

\newcommand{\Reals}{\mathbb{R}}
\newcommand{\Nats}{\mathbb{N}}

\newcommand{\cond}{\,\vert\,}

\DeclareMathOperator{\enc}{\mathrm{enc}}
\DeclareMathOperator{\dec}{\mathrm{dec}}
\DeclareMathOperator{\XSpace}{\mathcal{X}}
\DeclareMathOperator{\YSpace}{\mathcal{Y}}

\DeclareMathOperator{\ZSpace}{\mathcal{Z}}

\DeclareMathOperator{\Exp}{\mathbb{E}}

\DeclareMathOperator{\strings}{\{0, 1\}^*}
\DeclareMathOperator{\Prob}{\mathbb{P}}
\DeclareMathOperator{\Ent}{\mathnormal{H}}
\DeclareMathOperator{\DiffEnt}{\mathnormal{h}}
\DeclareMathOperator{\defeq}{\,\triangleq\,}
\DeclareMathOperator*{\argmin}{\mathrm{argmin}}

\DeclareMathOperator*{\esssup}{\mathrm{ess\,sup}}

\DeclareMathOperator{\Unif}{\mathrm{Unif}}
\DeclareMathOperator{\Bern}{\mathrm{Bern}}
\DeclareMathOperator{\Geom}{\mathrm{Geom}}
\DeclareMathOperator{\Exponential}{\mathrm{Exp}}

\DeclareMathOperator{\Oh}{\mathcal{O}}

\newcommand{\CrossEnt}{\Ent\infdivmi}

\DeclareSymbolFont{cyrletters}{OT2}{wncyr}{m}{n}
\DeclareMathSymbol{\Sha}{\mathalpha}{cyrletters}{"58}
\DeclareMathSymbol{\sha}{\mathalpha}{cyrletters}{"78}

\usepackage{thmtools}
\usepackage{thm-restate}

\usepackage{cleveref}

\crefname{thm}{theorem}{theorems}

\begin{document}
\onecolumn

%
\title{Functional Information, Optimal Channel Simulation and Lossy Source Coding}

\author{Gergely Flamich*, Spencer Hill*, Fady Alajaji and Tam{\'a}s Linder
\thanks{G.~Flamich is at Imperial College London, while S.~Hill, F.~Alajaji, and T.~Linder are with the Department of Mathematics and Statistics, Queen’s University, Kingston, Ontario, Canada. The authors acknowledge financial support from Imperial
College London through an Imperial College Research Fellowship grant
awarded to G.~Flamich. S.~Hill, F.~Alajaji, and T.~Linder acknowledge support from the Natural Sciences and Engineering Research Council of Canada.}
\thanks{This paper was presented in part at the 2026 IEEE International Symposium on Information Theory (Recent Results Session) and the 2026 IEEE Information Theory Workshop.}%
\thanks{* denotes equal contribution.}
}



\maketitle

\begin{abstract}
We characterize, within a constant number of bits, the minimum
achievable rate for several one-shot coding problems, such as
channel simulation and lossy source coding. 
The basis of our work
is a new information measure between random variables $X, Y$ that we
call the \textit{functional information} $\FI{X}{Y}$.  We show that
this quantity characterizes the minimum one-shot rate $R^*(X \to Y)$
of channel simulation, in which a sample $Y \sim P_{Y \mid X}$ is
communicated in the presence of common randomness:
$\FI{X}{Y} \leq R^*(X \to Y) < \FI{X}{Y} + 1.45$.  For the
achievability result, we construct an explicit code using the
Poisson functional representation.  
\par
Using these results, we develop
new bounds for one-shot variable-rate lossy source coding: given some distortion
measure $\Delta$, we use $\FI{X}{Y}$ to define the
\textit{functional rate-distortion function} $R_F$. Then, letting $R_O$ denote the one-shot operational rate-distortion function with private encoder randomness and distortion measure
$\Delta$, we make use of our channel simulation result to show
that $R_F(D) \leq R_O(D) < R_F(D) + 2.45$ at every positive distortion level
$D$.  
\par
Finally, we generalize our channel simulation result by
introducing the \textit{R{\'e}nyi $\alpha$-functional information}
$\RFI{\alpha}{X}{Y}$ and show that it characterizes the minimum
Arimoto-R\'enyi conditional entropy for channel simulation,  ${R^*_\alpha(X \to
Y)}$, associated with the Campbell cost: 
${\RFI{\alpha}{X}{Y} \leq R^*_{\alpha}(X \to Y) \leq
\RFI{\alpha}{X}{Y} + c_\alpha}$, where $c_\alpha$ is a constant depending on $\alpha \in (0, 1)$.
\end{abstract}


\section{Introduction} \label{sec:introduction}
\IEEEPARstart{C}{haracterizing} the minimum number of bits required to encode some data is a central aim of information theory.
Yet, despite some steady progress in recent years, the field still lacks tight nonasymptotic results for several fundamental problems.
Our paper makes progress in addressing this gap.
\par
Our foundational result is a tight nonasymptotic characterization of the best achievable rate for the problem known as channel simulation (relative entropy coding)~\cite{li2024channel, flamich2026data}.
Let $X$ and $Y$ be  random variables
taking values in the Polish spaces $\XSpace$ and $\YSpace$,
respectively, with joint distribution $P_{X,Y}$. Consider the
induced Markov kernel (channel) $P_{Y \mid X = x}$ and  the triplet
$(Z, \enc, \dec)$ consisting  of
\begin{itemize}
\item $Z \in \ZSpace$, a random variable independent of $X$ (denoted $Z\perp X$) called the \emph{common randomness}, which is shared by the encoder and decoder;
\item $\enc: \XSpace \times \ZSpace \to \strings$, an \emph{encoder} mapping an input and a realization of the common randomness to a binary string, with the induced code being uniquely decodable for each realization of the common randomness (where $\strings$ is the set of all finite-length binary strings);
\item $\dec : \strings \times \ZSpace \to \YSpace$, a \emph{decoder} mapping the encoder output and common randomness to a point $y \in \YSpace$. 
\end{itemize}
We say that $(Z, \enc, \dec)$ is a stochastic code for $P_{Y \mid X}$ if, for $P_X$-almost all inputs $x \in \XSpace$,
\begin{equation} \label{eq:stoch-code} 
\dec(\enc(x, Z), Z) \sim P_{Y \mid X = x}.
\end{equation}
When no confusion arises, we identify $Y$ with the code's output
$\dec(\enc(X, Z), Z)$ on input $X$. 
Letting~$\abs{s}$ denote the length of a binary string $s\in
\{0,1\}^*$, we are
interested in describing the minimum  expected code-length of any
stochastic code, 
\begin{align}
\label{eq:operational_rate}
L^*(X\to Y) \defeq \inf_{(Z, \enc, \dec)} \Exp[\abs{\enc(X, Z)}],
\end{align}
where the infimum is taken over all stochastic codes.

While we keep the above quantity as the main object of practical
interest, we can somewhat simplify the problem by instead considering
the minimum conditional entropy of the output of a stochastic code, 
\begin{align}
R^*(X \to Y) \defeq \inf_{(Z,\enc,\dec)}  \Ent(Y \mid Z).
\end{align}
By standard lossless source coding results, these quantities have the
simple relationship
\begin{equation} \label{eq: minentropycost} 
 R^*(X \to Y) \le
L^*(X\to Y) \le  R^*(X\to Y) +1. 
\end{equation} 
\par
The minimum achievable rate of stochastic codes was first
characterized for discrete channels in terms of the mutual information
$\MI{X}{Y}$ by Harsha et al.~\cite{harsha2010communication}, and later generalized to arbitrary channels by Li and El Gamal~\cite{li2018strong}.
General bounds on $R^*(X \to Y)$ were derived in the expanded version of~\cite{li2025discrete}, where it was shown that
\begin{align*}
\MI{X}{Y} \leq R^*(X \!\to \!Y)\!< \!\MI{X}{Y} \!+ \!\log(\MI{X}{Y} \!+\! 1)\! + \!2.74.
\end{align*}
While these results are  undoubtedly remarkable, there remains a logarithmic gap (i.e., a gap of order $\log(\MI{X}{Y}+1)$) between the upper and lower bounds, which has been shown to be necessary in general: 
there exist channels for which $R^*(X \to Y)$ is within a constant of the upper bound and others for which it is within a constant of the lower bound \cite{braverman2014public,li2018strong}.
\par
In light of the above fact, one suspects that a new information measure is needed to close this gap.
To this end, we propose the \textit{functional information}
$\FI{X}{Y}$ (see \Cref{def:I_F} in \Cref{sec:I_F})  and prove the following result, which forms the basis of all further results in the paper.
\begin{restatable}[]{theorem}{firatesandwich}
\label{thm:fi_rate_sandwich}
Let $X$ and $Y$ be jointly distributed random variables over an arbitrary Polish space.
Then,
\begin{align}
\FI{X}{Y} \leq R^*(X \to Y) \leq \FI{X}{Y} + \log e. \label{eq:main_result}
\end{align}   
\end{restatable}
\Cref{thm:fi_rate_sandwich} demonstrates that $\FI{X}{Y}$ is the ``correct'' information measure for the channel simulation problem. 
A special case of the lower bound in \eqref{eq:main_result} for discrete $Y$ has previously appeared in different notation as Proposition 1 in \cite{li2018strong} and was later proven for arbitrary $Y$ in \cite{flamich2024data,flamich2025redundancy}.
Here, we give a new, much-simplified proof of the general lower bound result (\Cref{thm:fi_lb}).
Moreover, by analyzing the equality condition, we fully characterize the conditions for so-called \emph{perfect functional representations} (see \Cref{prop:perf_fun_rep}), thus significantly extending the results of \cite{coban2026perfect}.
The upper bound in \eqref{eq:main_result} has only recently been shown to be achievable (within $\log e$ bits) for additive exchangeable noise channels with finite field input in \cite{zhao2025rejection}. 
However, contemporaneously, we independently developed a significantly more general achievability result for arbitrary channels in \cite{hill2026rejection} by providing an explicit code constructed using rejection sampling.
Unfortunately, the result in \cite{hill2026rejection} only holds under restrictive assumptions on the joint distribution of $X$ and $Y$.
In this paper, we construct a different code based on the Poisson
functional representation \cite{li2018strong} that completely
removes this restriction (see \Cref{thm:rtc_for_pfr} in \Cref{sec:upper_pfr}). 
This allows us to state the upper and lower bounds at the same level
of generality.\footnote{During the final preparation of this manuscript, Li~\cite{li2026optimal}
showed that the Poisson functional representation with proposal distribution $P_Y$ yields
$H(Y\mid Z)\le I_{\mathrm E}(X\to Y)+\log e$, where $I_{\mathrm E}(X\to Y)$ is the so-called {\em exponential} functional information, satisfying $I_{\mathrm E}(X\to Y)\le\FI{X}{Y}$.
For any dominating proposal distribution $Q_Y$, the proof of our 
Theorem~\ref{thm:rtc_for_pfr} gives the same upper bound on the
cross-entropy of our coding distribution for the minimizing sample index $K$ (see Remark~\ref{rem-exp-fi-li26} for more details).}
Two immediate consequences of \Cref{thm:fi_rate_sandwich} are (i)  a
mutual information-based upper bound with a tighter constant than that
in~\cite{li2025discrete} (see \Cref{cor:mi_based_one_shot_sandwich})
and (ii)  the recovery, using Corollary~\ref{cor:noredund}, of the results of \cite{sriramu2024optimal,flamich2025redundancy} on the second-order redundancy of asymptotic channel simulation (see \Cref{cor:chansim_asymptotic_redundancy}).
\par
Moreover, as the main application of \Cref{thm:fi_rate_sandwich}, we
characterize within a constant the best achievable rate in
variable-rate lossy
source coding for any finite blocklength in terms of the functional
information. For a fixed source distribution $P_X$, given a distortion measure $\Delta: \XSpace\times \YSpace
\to [0,\infty)$, we define the
\emph{functional rate-distortion function} $R_F(D)$ as
\begin{align}\label{eq:rfd}
R_F(D) \defeq \inf_{P_{Y \mid X}\, : \, \Exp[\Delta(X, Y)] \leq D} \FI{X}{Y}.   
\end{align}
Analogously, we define the variable-rate operational rate-distortion
function as
\begin{align}\label{eq:rod}
R_O(D) \defeq \inf_{\substack{P_{Y \mid X}:\, Y\ \text{countably-valued}\\ \Exp[\Delta(X,Y)]\le D}} H(Y).
\end{align}
Here the encoder uses private randomness to draw $Y$ according to
$P_{Y \mid X}$ and sends its index to the decoder. Following Li and El Gamal~\cite[Proof of Thm.~2]{li2018strong}, the
encoder can achieve $R_O(D)$ to arbitrary precision by randomizing
between two deterministic quantizers $Y=g_0(X)$ and $Y=g_1(X)$.
We prove the following result in \Cref{sec:rd_application}.

\begin{restatable}[]{theorem}{thmoperationalrd}
  \label{thm:one_shot_rd_fn}
Let $\XSpace$ and $\YSpace$ be Polish spaces. For any source
distribution $P_X$ and any $D>0$,
\begin{equation}
R_F(D) \le R_O(D) \le R_F(D) + \log e + 1. \label{eq:rdbound}
\end{equation}
\end{restatable} 

Now let $X^n =
(X_1,\ldots,X_n)$ and $Y^n=(Y_1,\ldots,Y_n)$ be $n$-blocks and
consider the average single-letter distortion measure  $\Delta_n(x^n,y^n) =\frac{1}{n}
\sum_{i=1}^n \Delta(x_i,y_i)$. If we replace  $\Delta$ with 
$\Delta_n$ and let   $R_{F,n}(D)$ and $ R_{O,n}(D)$
be the $n$th-order  versions of $R_F(D)$ and $R_O(D)$,  normalized by
$n$, then
Theorem~\ref{thm:one_shot_rd_fn} implies
\begin{equation} \label{eq:rf-normalized}
R_{F, n}(D) \le R_{O, n}(D) \le R_{F, n}(D) + \frac{\log e + 1}{n}.
\end{equation}

The proof of \Cref{thm:one_shot_rd_fn} uses
\Cref{thm:fi_rate_sandwich} and the two-quantizer construction.  Classically, the fundamental limits of
lossy compression were achievable only in the asymptotic
limit. Previous finite-length bounds or refined asymptotic analyses
(see \cite{zhou2023finite} for an overview) provide lower and upper
bounds with a logarithmic gap in the block-length of the encoded source,
or are valid only for
particular distortions and source distributions
\cite{Ordentlich2025TheVS}. A recent single-shot upper  bound
\cite{Kostina2026SingleShot} also leaves a logarithmic gap over the
rate-distortion function lower bound.  Our result, in terms
of the functional information, is tight within a constant and valid
for any source distribution, distortion measure, and blocklength. In
particular, we are not aware of any previous result that gives nonasymptotic lower and upper bounds on the optimal normalized  rate in lossy coding $n$-blocks under an expected distortion constraint with only a
$c/n$ gap for an absolute constant $c$  as in \eqref{eq:rf-normalized}. 

\par
Finally, in Section~\ref{sec:generalizations}  we define the R{\'e}nyi
generalization of the functional information $\RFI{\alpha}{X}{Y}$ for
$\alpha\in (0,1)$,
having the property  $\lim_{\alpha\to 1} \RFI{\alpha}{X}{Y} = \FI{X}{Y}$.
Then, we show that it too operationally characterizes the channel simulation problem under the so-called Campbell cost~\cite{campbell1965coding}.
Concretely, for $t>0$ let the Campbell (exponential) cost of the
stochastic code $(Z,\enc,\dec)$  be
\begin{equation}
  \label{eq:expcost}
L_t  \defeq \frac{1}{t} \log \left( \Exp \left[2^{t
      \abs{\enc(X,Z) }} \right] \right),
  \end{equation}
  and define the optimal Campbell cost over all stochastic codes  as
\begin{equation}
    \label{eq:optexpcost}
   L^*_t(X\to Y) \defeq \inf_{(Z,\enc,\dec)} L_t. 
\end{equation}
A related quantity is the minimum Arimoto-R\'enyi conditional entropy $H_\alpha^{\mathcal{A}}$
of order 
$\alpha\in (0,1)$ (defined in~\eqref{eq:arimotocre})~\cite{arimoto1977information} over all stochastic codes: 
\begin{align}\label{eq:optARent}
R^*_\alpha(X \to Y) \defeq \inf_{(Z,\enc,\dec)} 
  H_\alpha^{\mathcal{A}}(Y \mid Z).
\end{align}
It follows from the main result of~\cite{campbell1965coding} that
with $\alpha =\frac{1}{t+1}$,  these quantities are related as
\begin{equation}
  R^*_\alpha(X \to Y)\le  L^*_t(X\to Y)   \le R^*_\alpha(X \to Y) +1.
\end{equation}

Given the above, we prove the following result.

\begin{restatable}[]{theorem}{thmrenyigen} \label{thm:renyi_pfr} 
Let $X, Y$ be jointly distributed  random variables over  arbitrary
Polish spaces and let 
$\alpha \in (0, 1)$.
Then,
\begin{align}
\RFI{\alpha}{X}{Y} \leq R^*_\alpha(X \to Y) \leq \RFI{\alpha}{X}{Y} +
  \frac{1 - \alpha}{\alpha} + \frac{\log(1 / \alpha)}{1 - \alpha}. \label{eq:renyi_pfr_theorem}
\end{align}
\end{restatable}
As $\alpha \to 1$, the additive gap tends to $\log e$, recovering the
constant in \Cref{thm:fi_rate_sandwich}. We sharpen the channel simulation
bounds of~\cite{hill2025communication} by bounding $R^*_\alpha(X\to Y)$
above and below in terms of the same quantity, $\RFI{\alpha}{X}{Y}$,
with a gap depending only on $\alpha$. Fan and Wagner~\cite{fan2026exact}
independently obtained a related characterization for exact sampling
with a fixed target and proposal. Our result treats the full channel
simulation problem, in which a single code must simulate the conditional
law $P_{Y\mid X}(\, \cdot\cond x)$ for $P_X$-almost every $x$, with $x$
known only to the encoder.
\subsection{Organization}
Section~\ref{sec:preliminaries} introduces the channel simulation
divergence and functional information and establishes their basic
properties. Section~\ref{sec:perfect_functional} proves the one-shot
 lower bound in our main result and characterizes perfect functional
representations. Section~\ref{sec:rtc} establishes the upper bound of 
the main result via
the Poisson functional representation and derives consequences for
one-shot and asymptotic channel
simulation. Section~\ref{sec:rd_application} develops bounds for
one-shot and finite-blocklength lossy compression using the functional
rate-distortion function. Section~\ref{sec:generalizations} extends
the channel simulation bounds to R\'enyi functional information and
the Campbell cost. The appendices contain additional proofs and 
rejection-sampling constructions.

\subsection{Notation}
We denote the binary logarithm as $\log$ and the natural logarithm as $\ln$. 
For an event $A$, its indicator function is $\Ind{A}$.
Unless otherwise stated, all random variables are assumed to take values in an arbitrary Polish space.
Random variables are written in capital letters such as $X$. We write
$X \sim P_X$ to say that $X$ has probability measure $P_X$, and use
$P_X^{\otimes n}$ for the joint distribution of $n$ independent draws
from $P_X$. We will occasionally write $\text{Law}(X)$ to refer to the
distribution (law) of a random variable $X$. Random variable alphabets or spaces are denoted by calligraphic letters $X \in \XSpace$, $Y \in \YSpace$. 
We denote the expectation of $X$ and the conditional expectation of $X$ given $Y$ as $\Exp[X]$ and $\Exp[X \mid Y]$, respectively. We write $X \perp Y$ to indicate that $X$ and $Y$ are independent and use i.i.d. as shorthand for independent and identically distributed.
For two probability measures $P$ and $Q$ on the same measurable space, we write $P \ll Q$ to indicate that $P$ is absolutely continuous with respect to $Q$, and $\RND{P}{Q}$ denotes their Radon-Nikodym derivative. 
For discrete distributions $P$ and $Q$, we denote the Shannon entropy of $P$ as $\Ent(P)$ and write $\CrossEnt{P}{Q} \defeq \Exp_{X \sim P} [-\log Q(X)]$ for their cross-entropy.
For a distribution $P$ over the reals with density function $f$, we denote its differential entropy as $\DiffEnt(f)$ or $\DiffEnt(P)$. 
For a $Q$-measurable function $r$, we denote its $Q$-essential supremum as $\norm{r}_\infty$.
All information measures are in bits.

\section{Preliminaries} \label{sec:preliminaries}
\subsection{Channel Simulation Divergence}
\label{sec:intro_csd}
For two dependent random variables $X, Y$, the functional information of the channel $X \to Y$ is defined as a conditional channel simulation divergence ${\FI{X}{Y} \defeq \CSD{P_{X \mid Y}}{P_X \mid P_Y}}$. 
Thus, we begin the paper by first defining the channel simulation
divergence and establishing its basic properties. Throughout this
paper, we assume that $X$ and $Y$ are jointly distributed Polish space-valued random variables with finite mutual information
$\MI{X}{Y}<\infty$. This, in particular, implies that $P_{Y|X=x}\ll
P_Y$ for $P_X$-almost every (a.e.)~$x$. 
\par
\begin{figure}
    \centering
    \includegraphics[]{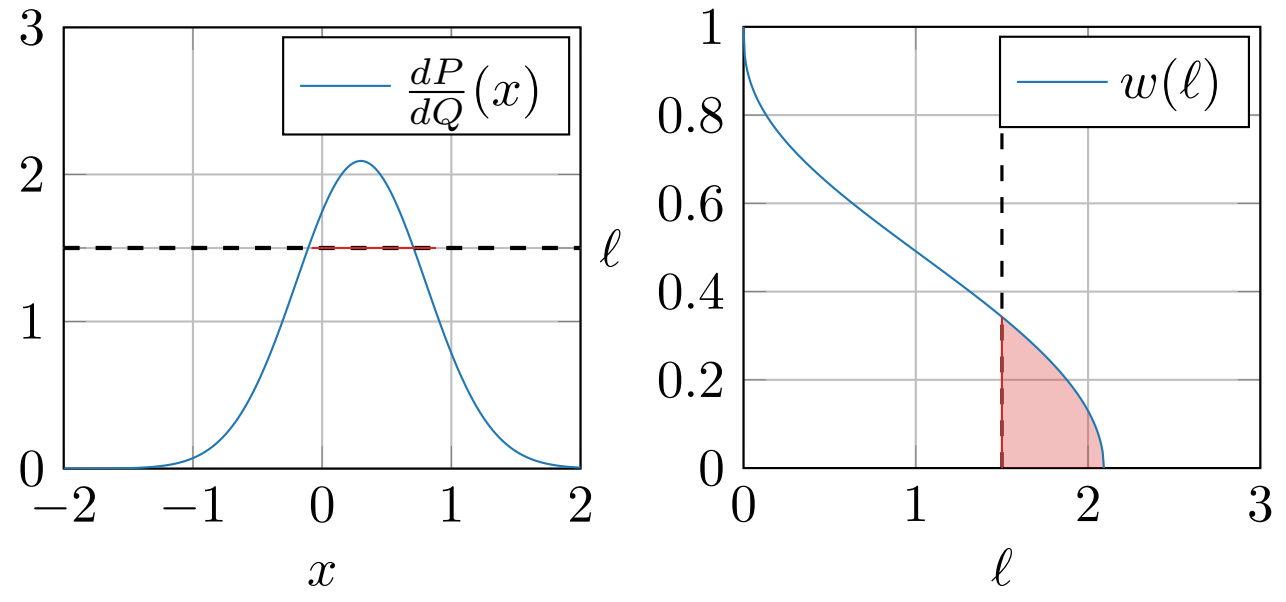}
    \caption{Visualization of the width function when $P = \Normal{0.3}{0.25}$ and $Q = \Normal{0}{1}$. \textbf{Left:} the density ratio $\RND{P}{Q}$; the value of the width function at $\ell = 1.5$ is the width of the \textcolor{tabred}{red} line. \textbf{Right:} the width function $w(\ell)$, with the value at $\ell = 1.5$ also marked as in the left plot.}
    \label{fig:width}
\end{figure}
\begin{definition}[Width function]
\label{def:width_fn}
Let $P \ll Q$ be probability measures over an arbitrary Polish space.
The width function $w : [0, \infty) \to [0, 1]$ of the pair $(P,Q)$ is defined as
\begin{align}
w(\ell) \defeq Q\left(\frac{dP}{dQ} \geq \ell\right).
\end{align}   
\end{definition}
\noindent
Note that $0 \leq w(\ell) \leq 1$ and
\begin{align}
\label{eq:width_fn_integrates_to_one}
\int_0^\infty w(\ell) \, d\ell = \Exp_{X\sim Q}\left[\frac{dP}{dQ}(X)\right] = 1.
\end{align}
Hence, $w$ is a valid probability density function (PDF) over the nonnegative reals; it is illustrated in \Cref{fig:width}. 
Now, we make the following definitions.
\begin{definition}[Channel simulation divergence, conditional channel simulation divergence]
Let $P, Q$ and $w$ be as in \Cref{def:width_fn}.
Then, the \textit{channel simulation divergence} of $P$ from $Q$ is defined as the differential entropy of $w$:
\begin{align}
\CSD{P}{Q} \defeq \DiffEnt(w) = - \int_0^\infty w(\ell) \log w(\ell) \, d\ell. 
\end{align}
Furthermore, for a probability measure $P_X$ and two Markov kernels
$P_{Y \mid X}$ and $Q_{Y \mid X}$ satisfying $P_{Y \mid X} \ll Q_{Y
\mid X}$ $P_X$-a.s., we first set $w_x(\ell) \defeq Q_{Y \mid X = x} \left( \RND{P_{Y \mid X}}{Q_{Y \mid X}} (Y \mid x) \ge \ell \right)$ to be the width function when $X = x$.
Then, we define
the \textit{conditional channel simulation divergence} is defined as
\begin{align}
\CSD{P_{Y \mid X}}{Q_{Y \mid X} \mid P_X} \defeq \Exp_{X \sim P_X}[\DiffEnt(w_X)].
\end{align}
\end{definition}
The channel simulation divergence was first defined in \cite{goc2024channel}, where the authors established some basic properties and used it to study channel simulation with so-called causal rejection samplers. One important property is its relation to the Kullback-Leibler (KL) divergence $\KLD{\cdot}{\cdot}$, which is summarized in the following proposition proved in~\cite[Lemma IV.1]{goc2024channel}.
\begin{proposition}[Relation to relative entropy]
\label{prop:csd_kl_sandwich_bound}
Let $P \ll Q$ be probability measures and let ${l(x) = x + \log(x + 1) + 1}$.
Then, 
\begin{align}
\KLD{P}{Q} \leq \CSD{P}{Q} \leq l(\KLD{P}{Q}).
\end{align}
\end{proposition}
The channel simulation divergence is also related to the recently defined \textit{layered entropy} $\lambda$ \cite{hegazy2022randomized} and \textit{discrete layered entropy} $\Lambda$~\cite{li2025discrete}.
For a more detailed study of the channel simulation divergence and its
properties, we refer the interested reader
to~\cite{flamich2024data}. One property of the channel simulation
divergence which is important to our results is that it is convex in
its first variable~\cite[Lemma IV.1]{goc2024channel}; a full proof is provided in Appendix~\ref{app:helper_proofs}.
\begin{proposition}[] \label{prop:csd_convexity}
For a fixed $Q$, the map $P \mapsto \CSD{P}{Q}$ is convex.
\end{proposition}
\subsection{Functional Information} \label{sec:I_F}
\begin{definition}[Functional Information] \label{def:I_F} 
Let $X, Y \sim P_{X, Y}$ be random variables.
Then, the \textit{functional information} of the channel $X \to Y$ is defined as
\begin{align}
\FI{X}{Y} \defeq \CSD{P_{X \mid Y}}{P_X \mid P_Y} = \Exp_Y[h(w_Y)],
\end{align}
where $w_y(\ell) = P_X \left( \RND{P_{Y \mid X}}{P_Y}(y \mid X) \ge
  \ell \right)$. 
More explicitly,
\[
\FI{X}{Y} = - \int_{\YSpace} \int_0^{\infty}  w_y(\ell) \log w_y(\ell) \, d\ell \, dP_Y(y). 
\]
\end{definition}
As suggested by the arrow notation, unlike the mutual information, the functional information is not generally symmetric in $X$ and $Y$. However, just as the channel simulation divergence is close to the KL divergence, $\FI{X}{Y}$ is always ``close'' to the mutual information, with equality if and only if the channel $X \to Y$ is \emph{singular} \cite{hill2026rejection}.
\begin{definition}[Singular channel \cite{sriramu2024optimal}]
\label{def:singularity} 
Let $X, Y \sim P_{X, Y}$. We call the channel $X \to Y$ singular if there exists a $P_Y$-measurable function $g(y)$ such that
\begin{align}
\frac{dP_{Y \mid X}}{dP_Y}(y \mid x) = g(y) \quad P_{X, Y}\text{-a.s.};
\end{align}
otherwise, we call the channel $X \to Y$ nonsingular.
\end{definition}
Singular channels (such as the binary erasure channel and the uniform dither channel) have been shown to admit relative entropy codes with sub-logarithmic asymptotic redundancy~\cite{sriramu2024optimal, flamich2025redundancy}. The following proposition shows that this is not an accident. Indeed, we will obtain a significantly stronger statement: constant redundancy over the mutual information is possible for all singular channels (see Corollary~\ref{cor:noredund}). 
\begin{proposition}[Relation to the mutual information]
\label{prop:fi_relation_to_mi}
Let $X, Y \sim P_{X, Y}$.
Then we have
\begin{align*}
\MI{X}{Y} \leq \FI{X}{Y} \leq \MI{X}{Y} + \log(\MI{X}{Y} + 1) + 1. 
\end{align*}
Furthermore, the lower bound is achieved if and only if the channel $X \to Y$ is singular.
\end{proposition}
\begin{proof}
    See Appendix~\ref{app:helper_proofs}.
\end{proof}
It is a simple corollary of Proposition~\ref{prop:fi_relation_to_mi} that in the special case where $X$ is discrete and $X \to Y$ is the trivial channel $Y = X$, we have $\FI{X}{Y} = H(X)$.
In addition to Proposition~\ref{prop:fi_relation_to_mi}, $\FI{X}{Y}$ and $\MI{X}{Y}$ also share some deeper
properties. 
For example, the functional information satisfies a data processing
inequality and analogous convexity/concavity properties, the proofs of which can be found in Appendix~\ref{app:helper_proofs}.
\begin{proposition}[Data processing inequality]
\label{prop:fi_dpi}
Let $X \rightarrow Y \rightarrow Z$ be a Markov chain. Then,
\begin{align}
\FI{X}{Y} \geq \FI{X}{Z}.
\end{align}
\end{proposition}
\begin{proposition}[Convexity/Concavity]
\label{prop:fi_convex_concave}
Consider the joint distribution $P_{X, Y}$ over $\XSpace \times \YSpace$.
Then:
\begin{itemize}
\item For fixed $P_{Y \mid X}$, the map $P_X \mapsto \FI{X}{Y}$ is concave.
\item For fixed $P_{X}$, the map $P_{Y \mid X} \mapsto \FI{X}{Y}$ is convex.
\end{itemize}
\end{proposition}
\section{One-shot Rate Lower Bound and Perfect Functional Representations} \label{sec:perfect_functional}
In this section, we give a new and simpler proof of the lower bound
$\FI{X}{Y} \le R^*(X \to Y)$ first shown
in~\cite{flamich2025redundancy}. A primary advantage of this proof is
that there is only \emph{one} inequality, permitting relatively simple
analysis of the equality condition. For $P_{X, Y}$ with finite
$\XSpace, \YSpace$, we give necessary and sufficient conditions on the
distribution for equality to be possible, generalizing the results of
\cite{coban2026perfect}.

\begin{definition}[Funcional Representation]  \label{def:funct-rep}
 Given a Polish space $\ZSpace$, let $Z \perp X$ be a $\ZSpace$-valued
 random variable and let $f: \XSpace\times\ZSpace\to\YSpace$ be measurable, with
 $Y = f(X, Z)$ and $P_{Y \mid Z = z}$ purely atomic for
 $P_Z$-a.e.~$z$. The pair $(Z,f)$ is called a functional
 representation for $P_{X,Y}$.  
\end{definition}

Note that for any pair of Polish space-valued random variables $X, Y$,
a functional representation $(Z, f)$ as just defined always
exists; for example, we may choose the Poisson functional
representation \cite{li2018strong} or the greedy Poisson rejection
sampler \cite{flamich2023greedy}. Also, any stochastic code, defined
in the introduction, gives rise to a functional representation  by
letting $f(X,Z) =\dec(\enc(X,Z),Z)$, and
any functional representation  can be converted to a stochastic code
by applying, conditioned on $Z=z$, a uniquely decodable code to the
(conditionally) discrete output $Y=f(X,Z)$. 

\begin{theorem}[Functional Information Lower Bound] \label{thm:fi_lb}
Let $X, Y \sim P_{X, Y}$ be random variables with $\MI{X}{Y} < \infty$. 
Let $Z \perp X$ be such that $Y = f(X, Z)$ for some measurable $f : \XSpace \times \ZSpace \to \YSpace$ and $P_{Y \mid Z= z}$ is purely atomic for $P_Z$-a.e.~$z$. Then,
\begin{equation}
\label{eq:fi_lower_bound}
\FI{X}{Y} \le R^*(X \to Y). 
\end{equation}
\end{theorem}
\begin{proof}
Let $Z \perp X$ be any random variable admitting a functional representation $Y = f(X, Z)$ for some measurable $f$ such that $P_{Y \mid Z = z}$ is purely atomic for $P_Z$-a.e.~$z$. To prove~\eqref{eq:fi_lower_bound}, it is sufficient to prove that $\FI{X}{Y} \le H(Y \mid Z)$. 
\par
For a fixed $y \in \YSpace$, define
\begin{equation*}
q_y(z) \defeq P_{Y \mid Z = z}(y) = \int_{\XSpace} \Ind{f(x, z) = y} \, dP_X(x).
\end{equation*}
Let $\nu_y$ be a regular conditional distribution of $Z$ given $ Y = y$. 
Note that $P_{Y \mid Z = z}(\qty{y : q_y(z) > 0}) = 1$ for $P_Z$-a.e.~$z$ implies $\Prob(q_Y(Z) > 0 \mid Z = z) = 1$ and hence $\Prob(q_Y(Z) > 0) = 1$. 
Hence, $\Prob(q_Y(Z) > 0 \mid Y = y) = 1$ for $P_Y$-a.e.~$y$.
Thus, we can choose versions of $\nu_y$ so that, for $P_Y$-a.e.~$y$, $q_y(z) > 0$ for $\nu_y$-a.e.~$z$. 
The conditional entropy is therefore
\begin{align} \label{eq:condentropy} 
H(Y \mid Z) \cdot \ln2 &= \Exp[-\ln q_Y(Z)] \nonumber \\ 
&= \int_{\YSpace} \int_{\ZSpace} - \ln q_y(z) \, d\nu_y(z) \, dP_{Y}(y).
\end{align}
To prove $\FI{X}{Y} \le H(Y \mid Z)$, it suffices to prove for $P_Y$-a.e.~$y$ that
\begin{equation}
\label{eq:cond_entropy_lower_bound}
\CSD{P_{X \mid Y = y}}{P_X} \cdot \ln 2 \le \int_{\ZSpace} - \ln q_y(z) \, d\nu_y(z). 
\end{equation}
We will achieve this using the representation 
\begin{equation}
    \CSD{P_{X \mid Y = y}}{P_X} \cdot \ln 2 = -1 + \int_0^1 \frac{1}{p} \int_0^p w_y^{-1}(u) \, du \, dp \label{eq:csd_representation}
\end{equation}
of~\cite[Lem. 3.1.2]{flamich2024data}, where 
\begin{equation}
    w_y^{-1}(u) \defeq \sup \qty{\ell \ge 0 : w_y(\ell) \ge u}
\end{equation}
is the generalized inverse of the width function $w_y$. \par
To begin, fix $y$ and write 
\begin{equation}
    r(y \mid x) = \RND{P_{Y \mid X}}{P_Y}(y \mid x). \label{eq:r_shorthand}
  \end{equation}
Note that $r(y\cond  x)=\frac{dP_{X\mid Y=y}}{dP_X}(x)$ for
$P_Y$-a.e.~$y$, $P_X$-a.e.~$x$ since $I(X;Y)<\infty$ implies
$P_{X,Y}\ll P_X\otimes P_Y$.

For every measurable $A \subseteq \XSpace$,
\begin{align*}
P_{X \mid Y = y}(A) &= \int_{\ZSpace} \Prob(X \in A \mid Y = y, Z= z) \, d\nu_y(z) \\
&= \int_{\ZSpace} \frac{\int_A \Ind{f(x, z) = y} \, dP_X(x)}{q_y(z)} \, d\nu_y(z), 
\end{align*}
where the second equality uses the independence of $Z$ and $X$ and the fact that, given $Z = z$, the event $\qty{Y = y}$ is $\qty{f(X, z) = y}$. Thus,
\begin{equation*}
r(y \mid x) = \int_{\ZSpace} \frac{\Ind{f(x, z) = y}}{q_y(z)} \, d\nu_y(z) \quad P_X\text{-a.s.}
\end{equation*}
Fix $p \in [0, 1]$ and consider any measurable $b_y : \XSpace \to [0,
1]$ with $P_{X}(b_y) = p$, where we used the convention $P(g)=\int
g\,dP$. 
Then, 
\begin{align}
P_{X \mid Y = y}(b_y) &=\int_{\XSpace} b_y(x) r(y \mid x) \, dP_X(x) \nonumber\\
&=  \int_{\ZSpace} \frac{\int b_y(x) \Ind{f(x, z) = y}\, dP_X(x)}{q_y(z)} \, d\nu_y(z) \nonumber \\
&\le \int_{\ZSpace} \frac{\min \qty{p, q_y(z)}}{q_y(z)} \, d\nu_y(z), \label{eq:supremum_over_b}
\end{align}
where the inequality follows because 
\begin{equation*}
\int_{\XSpace}  b_y(x) \Ind{f(x, z) = y} \, dP_X(x) \le \int_{\XSpace} b_y(x) \, dP_X(x) = p
\end{equation*}  
and 
\begin{align*}
\int_{\XSpace}  b_y(x) \Ind{f(x, z) = y} \, dP_X(x) 
&\le \int_{\XSpace} \Ind{f(x, z) = y} \, dP_X(x)
= q_y(z).
\end{align*}
Next, we note that
\begin{equation}
\sup_{\substack{0 \le b_y \le 1 \\  P_X(b_y) = p}} \int_{\XSpace} b_y r(y \mid x) \, dP_X(x) = \int_0^p w_y^{-1}(u) \, du.
\end{equation}
This representation is a standard result; see, e.g.,~\cite[Theorem 1.14]{lieb2001analysis}. We provide a proof along with the structure of the maximizing $b_y$ in Appendix~\ref{app:helper_proofs} (Lemma~\ref{lemma:bathtub}) for completeness. With this representation, taking the supremum over all admissible $b_y$ in~\eqref{eq:supremum_over_b} gives 
\begin{equation}
\int_0^p w_y^{-1}(u) \, du \le \int_{\ZSpace} \frac{\min \qty{p, q_y(z)}}{q_y(z)} \, d\nu_y(z). \label{eq:key_inequality_bathtub}
\end{equation}
We can multiply by $1/p$ and integrate to obtain
\begin{align*}
\int_0^1 \frac{1}{p} \int_0^p w_y^{-1}(u) \, du \, dp &\le \int_0^1 \frac{1}{p} \int_{\ZSpace} \frac{\min \qty{p, q_y(z)}}{q_y(z)} \, d\nu_y(z) \, dp \\
&=  \int_{\ZSpace} \int_0^1 \frac{\min \qty{p, q_y(z)}}{p q_y(z)} \, dp \, d\nu_y(z) \nonumber \\
&= 1 + \int_{\ZSpace} - \ln q_y(z) \, d\nu_y(z),
\end{align*}
where the first equality follows from Tonelli's theorem and the final equality follows from the computation
\begin{align*}
\int_0^1 \frac{\min \qty{p, q}}{pq} \, dp &= \int_0^q \frac{p}{pq} \, dp + \int_q^1 \frac{q}{pq} \, dp \\
&= 1 - \ln q.
\end{align*}
Substituting in~\eqref{eq:csd_representation} gives that
\begin{align*}
\CSD{P_{X \mid Y = y}}{P_X} \cdot \ln 2 
&\le -1 + \left(1 - \!\int_{\ZSpace} \ln q_y(z) \, d\nu_y(z) \right) \\
&= \int_{\ZSpace} - \ln q_y(z) \, d\nu_y(z).
\end{align*}
In view of the definition of $\FI{X}{Y}$ and  \eqref{eq:condentropy}, the result follows after integrating over $\YSpace$ with respect to
$d P_Y(y)$.
\end{proof}
One can observe that the only inequality in the proof of
Theorem~\ref{thm:fi_lb} is~\eqref{eq:supremum_over_b}.  Therefore, if
equality holds in~\eqref{eq:key_inequality_bathtub} for
a.e.~$p\in(0,1)$ and for $P_Y$-a.e.~$y$,  we have $\FI{X}{Y} = H(Y \mid Z)$, which we
call a \emph{perfect functional representation} (see
also~\cite{coban2026perfect} for a slightly different definition; see our \Cref{remark:perfect_func_rep} below).
\begin{definition}[Perfect Functional Representation]
For ${X, Y \sim P_{X, Y}}$, we call  a  functional representation  $(Z, f)$ \textit{perfect} if $H(Y \mid Z) = \FI{X}{Y}$. 
\end{definition}
When $\XSpace, \YSpace$ are finite, one can obtain necessary and
sufficient conditions on $P_{X, Y}$ for a perfect functional
representation to exist. Suppose that $\XSpace$ and $\YSpace$ are
finite with $\length{\XSpace} = n$ and $\length{\YSpace} = m$. Assume
without loss of generality that $P_X(x) > 0$ and $P_Y(y) > 0$ for all
$x \in \XSpace$ and $y \in \YSpace$. To ease the notation, we write
$P_{Y|X}(y\mid x)$ for $P_{Y|X}(\{y\}\mid x)$. 
\begin{definition}[Compatible ordering, admissible maps] 
We call an ordering $x_{y, 1}, x_{y, 2}, \ldots, x_{y, n}$ of $\XSpace$ $y$-\emph{compatible} if 
\begin{equation*}
P_{Y \mid X}(y \mid x_{y, i}) \ge P_{Y \mid X}(y \mid x_{y, j})
\end{equation*}
for all $i < j$. Given $y$-compatible orderings, one for each $y \in \YSpace$, we call a map $\varphi : \XSpace \to \YSpace$ \emph{admissible} if, for every $y \in \YSpace$, the preimage $\varphi^{-1}(y)$ satisfies
\begin{equation*}
\varphi^{-1}(y) = \qty{x_{y, 1}, \ldots, x_{y, k_y}}
\end{equation*}
for some $k_y \in \qty{0, 1, \ldots, n}$, with the convention that the
set is empty if $k_y = 0$. Equivalently, $\varphi$ is admissible if
each output cell $\qty{x : \varphi(x) = y}$ is an initial segment of
the fixed ordering compatible with $y$.

To any admissible  $\varphi$ uniquely corresponds a binary
(deterministic) $n\times m$ channel transition matrix
$M_{\varphi}=\{M_{\varphi}(x,y)\} $
such that $M_{\varphi}(x,y)=1$ if and only if $\varphi(x)=y$. We call
such a binary matrix admissible for the given ordering. 
\end{definition}
\begin{proposition}[Finite-Alphabet Perfect Representation] \label{prop:perf_fun_rep}
Assume that $\XSpace$ and $\YSpace$ are finite.
There exists a perfect functional representation if and only if there exist $y$-compatible orderings of $\XSpace$, one for each $y \in \YSpace$, and a probability mass function $\alpha$ on the set of admissible maps $\varphi : \XSpace \to \YSpace$ for these orderings, such that 
    \begin{equation}
        P_{Y \mid X}(y \mid x) = \sum_{\varphi : \varphi(x) = y} \alpha(\varphi) \label{eq:varphi_to_satisfy}
    \end{equation}
    for every $x \in \XSpace$, $y \in \YSpace$. Equivalently, the channel matrix $P_{Y \mid X}$ lies in the convex hull of the deterministic channel matrices associated with the maps admissible for some fixed choice of compatible orderings.
\end{proposition}
\begin{proof}
    The proof can be found in Appendix~\ref{app:perfect_functional}.
\end{proof}
\begin{remark}
\label{remark:perfect_func_rep}
In \cite{coban2026perfect}, the authors call a functional representation perfect if $H(Y \mid Z) = \MI{X}{Y}$. For discrete and singular $P_{X, Y}$, they give a necessary and sufficient condition in terms of four matrix operations for a perfect functional representation to exist. They also cast the problem of checking perfect representability as a linear program. Their linear program follows from our more general convex hull condition, after applying the additional constraint that $\FI{X}{Y} = \MI{X}{Y}$, i.e., that the channel $X \to Y$ is singular. 
\end{remark}
The conditions of Proposition~\ref{prop:perf_fun_rep} can be used to show that every binary-output channel admits a perfect representation, as was first shown in~\cite[Proposition 1]{li2018strong} and achieved by the Poisson functional representation.

\begin{proposition}[Binary-Output Perfect Functional Representation] \label{prop:bin_perf_rep}
Assume $\YSpace = \qty{0, 1}$ and let $Z \sim \Unif(0, 1)$ be independent of $X$. 
Let $f : \XSpace \times \ZSpace \to \YSpace$ be given by $f(x, z) = \Ind{P_{Y \mid X}(1 \mid x) \ge z}$. 
Then  $Y = f(X, Z)$ satisfies 
$(X,Y) \sim P_{X,Y}$ and  $H(Y \mid Z) = \FI{X}{Y}$. 
\end{proposition}
\begin{proof}
The proofs of correctness and optimality, both of which follow by computation, are deferred to Appendix~\ref{app:perfect_functional}. 
\end{proof}
\begin{remark}
  For the representation in
  Proposition~\ref{prop:bin_perf_rep}, the conditional distribution of
  $Y$ given $Z=z$ is determined by
\begin{equation*} 
P_{Y \mid Z}(1 \mid z) = P_X \left(P_{Y \mid X} (1 \mid X) \ge z \right) = w_1 \left(\frac{z}{P_{Y}(1)} \right).
\end{equation*}
One can show that Huffman coding for this distribution minimizes the expected code length among
  all stochastic codes, with the convention that no bits are sent when
  $Z$ determines $Y$.
\end{remark}

The conditions in Proposition~\ref{prop:perf_fun_rep} are relatively restrictive, and are not satisfied by every channel.
Below, we give an example of a discrete channel for which no perfect functional representation exists; the proof is in Appendix~\ref{app:perfect_functional}.
\begin{proposition}
\label{prop:no_perfect_fi}
Let $\XSpace = \YSpace = \qty{1, 2, 3}$ and define 
\begin{equation*}
P_{Y \mid X} = 
\begin{bmatrix}
1-\epsilon & 0 & \epsilon \\
\epsilon & 1-\epsilon & 0 \\
0 & \epsilon & 1-\epsilon
\end{bmatrix},
\end{equation*}
where $1/3 < \epsilon < 1/2$. Then, there does not exist a perfect functional representation for this channel. 
\end{proposition}
\section{One-Shot Rate Upper Bound via Poisson Functional Representation} \label{sec:rtc}
Having seen that perfect functional representations are not always possible, a natural question is whether there is a general algorithm that achieves a cost within a constant of the functional information lower bound. 
In this section, we answer this question in the positive by constructing an explicit code using the Poisson functional representation. 
\par
In Appendix~\ref{app:generalised_ring_toss}, we also present a generalization of our previous rejection
sampling-based code in \cite{hill2026rejection}, called the generalized ring toss code.  This code is of
independent interest, as it turns out that rejection sampling is 
optimal (within a constant gap) under any admissible dominating distribution of the proposed
samples in the common randomness, and that the choice which minimizes
the runtime is a Shtarkov distribution \cite{shtar1987universal}.
\subsection{Poisson Functional Representation} \label{sec:upper_pfr}
In \cite{hill2026rejection}, we constructed a rejection-sampling based
code, called the ring toss code, that achieves a cost within a
constant of the functional information lower bound. A limitation of
this rejection sampling algorithm is that it requires a uniformly
bounded density ratio $\frac{dP_{Y|X}}{dP_Y}$. Although our extension to an
arbitrary proposal distribution relaxes this requirement to the
channel having a finite Shtarkov integral (see Appendix~\ref{app:generalised_ring_toss}), this still excludes
many channels of practical interest such as the additive white
Gaussian noise channel.

In pursuit of removing this
restriction, we study how to apply the insights of the ring toss code
to the Poisson functional representation, which is a sampling
algorithm for general channels~\cite{li2018strong}. In particular, we
look to make use of  the common randomness to design a better coding
distribution for the minimizing sample index $K$ than the Zipf
distribution used to establish the fundamental upper bound in \cite{li2024channel}. For random variables
$X, Y \sim P_{X, Y}$, recall the Radon-Nikodym derivative
$r(y \cond x) = \RND{P_{Y \mid X}}{P_Y}(y\cond  x)$. The Poisson
functional representation sets the common randomness
$Z = \qty{(Y_i, T_i)}_{i \ge 1}$ to be a Poisson point process with
intensity measure $P_Y \otimes \lambda$; i.e., $Y_i \iid P_Y$ and $\{T_i\}_{i\ge1}$
is a rate-one Poisson process. On input $X = x$, it returns the index
$K$ given by
\begin{equation}
    K = \argmin_{i \ge 1} \frac{T_i}{r(Y_i \cond  x)}.
\end{equation}
Equivalently, the Poisson functional representation returns the index
of the first arrival time of the transformed Poisson process $\{T_i / r(Y_i \cond x)\}_{i\ge1}$. It follows from the theory of Poisson processes that $P_{Y_K\mid X=x}=P_{Y\mid X=x}$ for $P_X$-a.e.~$x$. Moreover,~\cite{li2018strong} showed that
\begin{equation}
    H(K) < \MI{X}{Y} + \log (\MI{X}{Y} + 1) + 4.
\end{equation}
The main result of this paper is that the Poisson functional
representation is an optimal stochastic code within a constant gap. Additionally, like
rejection sampling, the Poisson functional representation is also constant-gap
optimal using \emph{any dominating proposal distribution} $\qty{Y_i}_{i \ge 1} \iid Q_Y$, not just the marginal $P_Y$. 
\begin{theorem} \label{thm:rtc_for_pfr}
    Let $X, Y \sim P_{X, Y}$ be random variables with $\MI{X}{Y} <
    \infty$. Let $Z = \qty{(Y_i, T_i)}$ be a Poisson point process independent of $X$
    with mean measure $Q_Y \otimes \lambda$, where $P_{Y \mid X = x}
    \ll Q_Y$ for $P_X$-a.e.~$x$. Define $r_Q(y \cond x) \defeq
    r(y \cond x) \RND{P_Y}{Q_Y}(y)$ (with value zero where the latter derivative vanishes) and, for input $X = x$, define 
    \begin{equation}
        K \defeq \argmin_{i \ge 1} \frac{T_i}{r_Q(Y_i \cond x)},\qquad t/0\defeq+\infty\quad, t>0.
    \end{equation}
    Then $P_{Y_K\mid X=x}=P_{Y\mid X=x}$ for $P_X$-a.e.~$x$ and 
    \begin{equation}
        H(Y_K \mid  Z) \le \FI{X}{Y} + \log e.
    \end{equation}
    Moreover, this upper bound is achievable within one bit by
    encoding $K$ conditioned on $Z$ using  the Shannon-Fano code  
    for the distribution
    \begin{equation} \label{eq:qprior} 
        Q_{K \mid Z}(k\cond z) \defeq \frac{a_{y_k}^Q(t_k)}{\zeta(z)},
    \end{equation}
    where, for $t > 0$,
    \begin{equation}
        a_y^Q(t) \defeq \int_{\XSpace} \exp\left( - \frac{t}{r_Q(y
            \cond x)} \right) \, dP_X(x), \label{eq:a_y_def}
    \end{equation}
    and $\zeta(z) = \sum_{k=1}^\infty a_{y_k}^Q(t_k)$ is a normalizing factor.
  \end{theorem}
\begin{proof}
For fixed $x$, discard points with $r_Q(y \cond x)=0$ and set
$s=t/r_Q(y \cond x)$. The transformed process has intensity measure $P_{Y\mid X=x}\otimes\lambda$ and its first arrival time $\theta_Q$ is distributed as $\theta_Q \sim \Exponential(1)$, independently of its first mark,
which has distribution $P_{Y\mid X=x}$; see \cite[Appendix~A]{li2018strong}. This establishes the correctness of the Poisson functional representation under arbitrary proposal $Q_Y$. Thus,
\begin{equation}\label{eq:independence}
(X,Y_K)\sim P_{X,Y}, \; \theta_Q\sim\Exponential(1), \text{ and }
\theta_Q \perp (X,Y_K).
\end{equation}
We next show that $H(Y_K \mid Z) \le \FI{X}{Y} + \log e$  by bounding the entropy as
\begin{align}
    H(Y_K\mid Z) &\le \CrossEnt{P_{K \mid Z}}{Q_{K \mid Z}} \nonumber \\
    &= \Exp[- \log a_{Y_K}^Q(T_K)] + \Exp[\log \zeta(Z)]. \label{eq:crossentropytobound}
\end{align}
The finiteness of both expectations will be verified below.
For the second term in~\eqref{eq:crossentropytobound}, note that, by Campbell's formula~\cite[Proposition 2.7]{last2018lectures},
\begin{align}
    \Exp[\zeta(Z)] &= \int_{\YSpace} \int_0^\infty a_y^Q(t) \, dt \, dQ_Y(y) \\
    &= \int_{\YSpace} \int_{\XSpace} r_Q(y \cond x) \, dP_X(x) \, dQ_Y(y) \\
    &= \int_{\YSpace} \RND{P_Y}{Q_Y}(y) \, dQ_Y(y) = 1.
\end{align}
Therefore, $\zeta(Z)$ is almost surely finite. Also,
$\zeta(Z) \ge a_{Y_K}^Q(T_K)>0$ almost surely, so $Q_{K \mid Z}$ is a well-defined probability distribution. Moreover, by Jensen's inequality, 
\begin{equation}
    \Exp[\log \zeta(Z)] \le \log \Exp[\zeta(Z)] = 0.
\end{equation}
To bound the first term in~\eqref{eq:crossentropytobound}, write $R_Q
= r_Q(Y_K \cond X)$ and $R = r(Y_K \cond X)$, so that $T_K = R_Q \theta_Q$.
For $t>0$, define $a_y(t) \defeq \int_{\XSpace} \exp(-t/r(y \cond x))\,dP_X(x)$,
with the same convention when $r(y \cond x)=0$. Since
$0<\RND{P_Y}{Q_Y}(Y_K)<\infty$ almost surely, we have
\begin{align}
    a_{Y_K}^Q(T_K) &= \int_{\XSpace} \exp \left( -\frac{T_K}{r_Q(Y_K
                     \cond x)} \right) \, dP_X(x) \\
    &= \int_{\XSpace} \exp \left( - \frac{R\theta_Q
      \RND{P_Y}{Q_Y}(Y_K)}{r(Y_K \cond x) \RND{P_Y}{Q_Y}(Y_K)} \right) \, dP_X(x) \\
    &= \int_{\XSpace} \exp \left( -\frac{R\theta_Q}{r(Y_K \cond x)} \right)\, dP_X(x) = a_{Y_K}(R\theta_Q). 
\end{align}
Therefore,
\begin{equation}
    \Exp[-\log a_{Y_K}^Q(T_K)] = \Exp[- \log a_{Y_K}(R\theta_Q)].
\end{equation}
Using $dP_{X\mid Y=y}(x)=r(y \cond x) dP_X(x)$ gives, for $P_Y$-a.e.~$y$ and $t>0$,
\begin{equation}\label{eq:T-density}
f_{R\theta_Q\mid Y_K=y}(t)
=\int_{\{r(y \cond x)>0\}}\frac{e^{-t/r(y \cond x)}}{r(y \cond  x)}
\,P_{X\mid Y=y}(dx)
=a_y(t).
\end{equation}
Therefore, 
\begin{equation}
    \Exp[- \log a_{Y_K}(R \theta_Q)] = h(R \theta_Q \mid Y_K).
\end{equation}
The remainder of the proof will be upper-bounding this differential entropy. Define independent variables $U\sim\Unif(0,1)$ and
$V\sim\text{Gamma}(2,1)$, independent of $(X,Z)$, and set
\begin{equation}\label{eq:AB}
A=RU,\qquad B=AV=RUV.
\end{equation}
Here $V$ has density $f_V(v)=ve^{-v}\Ind{v>0}$.
We obtain, for $a>0$, that
\begin{align}\label{eq:A-density}
f_{A\mid Y_K=y}(a)
&=\int_{\{r(y \cond x)>0\}}\frac{\Ind{0<a\leq r(y \cond x)}}{r(y \cond
                     x)}
\,P_{X\mid Y=y}(dx) \nonumber \\
&=w_y(a). \nonumber 
\end{align}
Hence,
\begin{equation}\label{eq:hA}
h(A\mid Y_K)=\FI{X}{Y}.
\end{equation}
Moreover,
\begin{equation}\label{eq:beta-gamma}
f_{UV}(s)=\int_s^\infty\frac1v\,ve^{-v}\,dv=e^{-s},\qquad s>0,
\end{equation}
so $UV\sim\Exponential(1)$ and $(B,Y_K)$ has the same distribution as
$(R\theta_Q,Y_K)$.

Conditioned on $Y_K$, $V\to UV\to B$ is a Markov chain. The data processing inequality, along with the fact that $(U,V)$ is independent of $Y_K$, gives
\begin{equation}\label{eq:DPI}
I(V;B\mid Y_K)\leq I(V;UV\mid Y_K)=I(V;UV).
\end{equation}
Then, the scaling property of differential entropy gives
\[
h(B\mid V, Y_K)=h(A\mid Y_K)+\Exp[\log V],
 h(UV\mid V)=\Exp[\log V].
\]
Since $\FI{X}{Y}<\infty$ by Proposition~\ref{prop:fi_relation_to_mi}
and $\Exp[|\log V|]<\infty$, the preceding identities and~\eqref{eq:DPI}
justify the following entropy calculation:
\begin{equation}\label{eq:entropy-bound}
\begin{aligned}
h(R \theta_Q \mid Y_K)
&=h(B\mid Y_K)\\
&=h(B\mid V,Y_K)+I(V;B\mid Y_K)\\
&=h(A\mid Y_K)+\Exp[\log V]+I(V;B\mid Y_K)\\
&\leq h(A\mid Y_K)+\Exp[\log V]+I(V;UV)\\
&=h(A\mid Y_K)+h(UV)\\
&=\FI{X}{Y}+\log(e),
\end{aligned}
\end{equation}
where the last equality follows since $UV\sim\Exponential(1)$.
This proves that $\Exp[-\log a_{Y_K}^Q(T_K)]$ is finite.
Since $0<a_{Y_K}^Q(T_K)\leq 1$ and $\zeta(Z)\geq a_{Y_K}^Q(T_K)$,
the negative part of $\log\zeta(Z)$ is bounded by $-\log a_{Y_K}^Q(T_K)$;
its positive part is integrable because $\Exp[\zeta(Z)]=1$.
Thus,~\eqref{eq:crossentropytobound} is justified and gives the claimed entropy bound.
Finally, a Shannon code for $Q_{K\mid Z}$ assigns lengths
$\lceil-\log Q_{K\mid Z}(k\cond z)\rceil$ on its support, so its expected length
is less than $\CrossEnt{P_{K\mid Z}}{Q_{K\mid Z}}+1\leq\FI{X}{Y}+\log e+1$.
\end{proof}
Our coding distribution $Q_{K|Z}$  for the Poisson functional representation also has an intuitive explanation. For fixed $X =x$, the density ratio of the returned point $(Y_K, T_K)$ with respect to the proposal $Q_Y \otimes \lambda$ is
\begin{equation}
    \RND{P_{Y_K, T_K\mid X=x}}{Q_Y \otimes \lambda} (y, t) = \exp \left( -
      \frac{t}{r_Q(y \cond  x)} \right).
\end{equation}
Intuitively, a point $(y, t)$ will be returned if it is the minimum
value of the transformed process, which is equivalent to no point $T_i
/ r_Q(Y_i \cond x)$ falling into the interval $[0, t / r_Q(y \cond
x))$. Treating the remaining points as an independent, unobserved
Poisson process, this probability is exactly $\exp (-t / r_Q(y \cond x))$. Integrating this density over all possible values of $X$, we obtain 
\begin{align}
     \int_{\XSpace} \exp \left( - \frac{T_k}{r_Q(Y_k \cond x)} \right) \, dP_X(x)
    &= a_{Y_k}^Q(T_k). \label{eq:a_y_density} 
\end{align}
After normalizing to be a probability distribution, this is exactly $Q_{K \mid Z}$. 

\begin{remark}[Computational Complexity]
  Following \cite{hill2026rejection}, we call the encoding of $K$
  using the conditional coding distribution  $Q_{K \mid Z}$, defined in
  \eqref{eq:qprior}, the
  \emph{ring toss code}. The ring toss code for the Poisson
  functional representation removes the requirement in
  \cite{hill2026rejection} that
  the density ratio 
  $dP_{Y|X} / dP_Y$ be bounded while being optimal within $\log e$ bits. However, this comes at a price: the distribution
  $Q_{K \mid Z}$ is likely quite difficult to use, requiring computing
  an infinite sum. In light of this limitation, we view the coding
  distribution $Q_{K \mid Z}$ as primarily a theoretical tool for
  bounding $H(Y \mid Z)$ rather than a practical coding
  distribution. \par Moreover, the Poisson functional representation
  is a noncausal~\cite{liu2018rejection} algorithm, meaning it cannot
  be implemented on a computer. There exist implementable variants
  such as $A^*$ coding~\cite{flamich2022fast}, but they require
  that \begin{equation} \inf_{Q_Y} \Exp_X \left[ \esssup_{y\sim Q_Y} \RND{P_{Y \mid
          X}}{Q_Y} (y \cond X) \right] < \infty.
    \end{equation}
    We expect that the insights of this work can be used to design
    practical coding distributions for general channels with unbounded
    Shtarkov integral. In Appendix~\ref{app:gprs}, we exhibit an
    encoding scheme of the index returned by greedy Poisson rejection
    sampling~\cite{flamich2023greedy}, a causal algorithm for general
    channels, which is optimal within $2\log e + 2$ bits.
\end{remark}
\begin{remark}\label{rem-exp-fi-li26}
    During the final preparation of this manuscript, Li~\cite{li2026optimal} independently established the constant-gap optimality of the Poisson functional representation for channel simulation. Specifically, Li introduced a family of generalized functional information measures, of which the exponential functional information $I_{\mathrm E}(X\to Y)$ is a special case, and proved that the Poisson functional representation with proposal distribution $P_Y$ and common randomness $Z$
satisfies
\[
    H(Y\mid Z)\le I_{\mathrm E}(X\to Y)+\log e.
\]
In Li's family of generalized functional information measures, a uniform
auxiliary variable on $(0,1)$ recovers the likelihood-ratio tail formula
for $\FI{X}{Y}$, while an exponential
auxiliary variable of rate $e$ gives $I_{\mathrm E}(X\to Y)$. It was
also proved in~\cite{li2026optimal}  that $I_{\mathrm E}(X\to Y)\le\FI{X}{Y}$. Together with the
converse $\FI{X}{Y}\le R^*(X\to Y)$ of Flamich, Sriramu, and
Wagner~\cite{flamich2025redundancy}, this shows that the conditional
entropy of the Poisson functional representation is within $\log e$ bits of optimal. 

We draw the following differences between our result and that of~\cite{li2026optimal}. First, Li~\cite{li2026optimal} considers the Poisson functional representation with the marginal distribution $P_Y$ as its proposal distribution, whereas our result applies to an arbitrary dominating proposal distribution $Q_Y$. This additional flexibility allows the proposal distribution to be chosen independently of the channel output marginal. Second, our proof is constructive: we give an explicit coding distribution $Q_{K \mid Z}$ for the index~$K$, conditioned on the common randomness $Z$. The proof in~\cite{li2026optimal} considers $P_{K\mid Z}(k\mid z)$ as the $P_X$-mass of the set
of inputs for which point $K=k$, in contrast, each unnormalized weight in~\eqref{eq:qprior} depends only on
the corresponding point $(y_k,t_k)$, so our code avoids computing these
set probabilities.  

\end{remark}
\subsection{Corollaries of the Upper Bound}
In this section, we collect some immediate corollaries of our main results. While we state the results for the Poisson functional representation, which has fewer assumptions, for any channel with bounded Shtarkov integral the same results can be achieved using rejection sampling. \par
Theorem~\ref{thm:rtc_for_pfr}, along with the lower bound of
Theorem~\ref{thm:fi_lb}, gives a  characterization of the one-shot
rate of channel simulation that is complete up to a constant. 
\begin{manualthm}{thm:fi_rate_sandwich}
    Let $X, Y$ be jointly distributed random variables over an arbitrary Polish space. Then,
    \begin{equation*}
        \FI{X}{Y} \le R^*(X \to Y) \le \FI{X}{Y} + \log e.
    \end{equation*}
\end{manualthm}
The cost of channel simulation can also be written in terms of the mutual information, improving the constant from 2.74 to 2.45 in the bound of~\cite{li2025discrete}.
\begin{corollary}
\label{cor:mi_based_one_shot_sandwich}
For any $X, Y \sim P_{X, Y}$ with $\MI{X}{Y} < \infty$,
\begin{align*}
R^*(X \to Y)
&\leq \MI{X}{Y} + \log (\MI{X}{Y} + 1)  + 1 + \log e  \\
&<\MI{X}{Y} + \log (\MI{X}{Y} + 1) + 2.45.
\end{align*}
\end{corollary}
\begin{proof}
The result follows immediately by applying the upper bound in Proposition~\ref{prop:fi_relation_to_mi} to the result of Theorem~\ref{thm:rtc_for_pfr} and invoking Jensen's inequality. 
\end{proof}
Singular channels have been previously shown to admit relative entropy codes with sub-logarithmic redundancy in the asymptotic case~\cite{sriramu2024optimal, flamich2026singular}. 
Using Theorem~\ref{thm:rtc_for_pfr}, we immediately obtain a significantly stronger statement: all singular channels achieve a constant one-shot redundancy over the mutual information. 
\begin{corollary} \label{cor:noredund}
Let $X, Y \sim P_{X, Y}$ with $X \to Y$ singular. Then,
\begin{equation}
R^*(X \to Y) \leq \MI{X}{Y} + \log e.
\end{equation}
\end{corollary}
Together, Theorem~\ref{thm:rtc_for_pfr} and Corollary~\ref{cor:noredund} immediately recover previous results on the asymptotic rate of relative entropy coding ~\cite{sriramu2024optimal,flamich2025redundancy}. 
Concretely, given $P_{X, Y}$, consider the problem of simulating $n$
independent copies of the channel $X \to Y$, i.e., letting $(X^n,Y^n) 
\sim (P_{X,Y})^{\otimes n}$, letting the proposal distribution be $P_{Y}^{\otimes n}$ and, upon input $X^n \sim P_{X}^{\otimes n}$, sending a binary message so the decoder can generate a sample $Y^n \sim P_{Y^n \mid X^n = x^n}$. In this problem, of interest is the best achievable rate per dimension
\begin{equation}
    R_n \defeq \tfrac{1}{n} \inf\nolimits_{(Z, \enc,\dec)} \Exp[ \lvert \enc(X^n, Z) \rvert].
\end{equation}
The distribution $P_{X^n, Y^n}$ satisfies ${\MI{X^n}{Y^n} = n
  \MI{X}{Y}}$, and hence ${R_n \to \MI{X}{Y}}$ as ${n \to
  \infty}$. Given the fundamental lower bound ${\Ent(Y \mid Z) \geq
  \MI{X}{Y}}$, this is the optimal limit. 
In light of the logarithmic gap over the mutual information in the one-shot case,~\cite{sriramu2024optimal,flamich2025redundancy} study the logarithmic redundancy 
\begin{equation}
    R^{\text{log}} \defeq \lim_{n \to \infty} \frac{n (R_n - \MI{X}{Y})}{\log n}.
\end{equation}
Whenever the limit exists, the preceding bounds give $0 \leq R^{\text{log}} \leq 1$. As a corollary to Theorem~\ref{thm:rtc_for_pfr}, we obtain the same characterization of $R^{\text{log}}$ as~\cite{flamich2025redundancy}:
\begin{corollary}
\label{cor:chansim_asymptotic_redundancy}
    Under the asymptotic setup previously described, for any $X, Y \sim P_{X, Y}$ whose information density has a finite second moment, 
    \begin{equation}
        R^{\textup{log}} = \begin{cases}
            0, &X \to Y \text{ singular} \\
            1/2 &X \to Y \text{ nonsingular}. 
        \end{cases}
    \end{equation}
\end{corollary}
\begin{proof}
The singular case is immediate from Corollary~\ref{cor:noredund}. 
For the nonsingular case, we use a result from the proof of Theorem 3.1.6 in \cite{flamich2025redundancy}, which shows that nonsingular channels whose information density has a finite second moment satisfy
\begin{equation}
\lim_{n \to \infty} \frac{\FI{X^n}{Y^n} - \MI{X^n}{Y^n}}{\log n} = \frac{1}{2}.
\end{equation}
Combining the above result with \Cref{thm:rtc_for_pfr} completes the proof.
\end{proof}
Our construction has the advantage over~\cite{sriramu2024optimal, flamich2026singular} of being significantly simpler and one-shot optimal. 
In contrast, in the singular case, the achievability algorithms of both~\cite{sriramu2024optimal} and~\cite{flamich2026singular} incur an $O(\log \log n)$ penalty in total length over $n\MI{X}{Y}$. 
\section{One-shot and finite-blocklength lossy compression} 
\label{sec:rd_application}
An immediate and quite interesting application of our results is the
tight characterization of finite blocklength lossy source coding. For
a source $P_X$ and (measurable) distortion measure $\Delta : \XSpace \times \YSpace \to [0, \infty)$, the classical rate-distortion function 
\begin{equation}
    R(D) = \inf_{P_{Y \mid X}\, :\, \Exp[\Delta(X, Y)] \le D }
    \MI{X}{Y} 
\end{equation}
is achievable in general by letting the blocklength grow to $\infty$. Using our results, we are able to tightly characterize the achievable-rate region for any blocklength $n$. Our proofs rely on the channel simulation results proved earlier. Our characterization relies on a measure we call the \emph{functional rate-distortion function}. We formally state the definitions in~\eqref{eq:rfd} and~\eqref{eq:rod}.
\begin{definition}[Functional rate-distortion function]
\label{def:frdf}
Fix a source $P_X$ and a distortion measure $\Delta$. We make the
usual assumption that $\min_{y\in \YSpace}  \Delta(x,y) =0$ for all $x\in
\XSpace$ and that $\Exp[\Delta(X,y^*)] < \infty$ for some $y^*\in
\YSpace$. Then, the functional rate-distortion function is defined as
\begin{align} \label{eq:def:frdf}
R_F(D) \defeq \inf_{P_{Y \mid X}\, :\,  \Exp[\Delta(X, Y)] \leq D } \FI{X}{Y}.
\end{align}   
\end{definition}
Note that $R_F(D)$ is nonincreasing, and by
\Cref{prop:fi_convex_concave}, it is also convex in $D$ on $(0,
\infty)$.

We are interested in bounding the rate of variable-length lossy source
coding of $X$ subject to a distortion constraint. As in the introduction,
the operational rate-distortion function is
\begin{equation}
  \label{eq:onerddef}
  R_O(D) \defeq \inf_{\substack{P_{Y \mid X}:\, Y\ \text{countably-valued}\\ \Exp[\Delta(X,Y)]\le D}} H(Y).
\end{equation}
The encoder may use private randomness to generate the reproduction $Y$.
It sends a uniquely decodable description of $Y$ to the decoder.
We have the following result.
\thmoperationalrd*
For the upper bound, the encoder draws a single Bernoulli variable $B$
independently of $X$ and encodes the pair $(B,g_B(X))$, as
in~\cite[Proof of Thm.~2]{li2018strong}. The selector contributes at most
one bit to the joint entropy.

Theorem~\ref{thm:one_shot_rd_fn} has an immediate consequence for
finite-blocklength lossy source coding. Let $X^n =
(X_1,\ldots,X_n)$ be an $n$-block of $\XSpace$-valued random
variables and let the reproduction $n$-block
be $Y^n=(Y_1,\ldots,Y_n)$. Define $\Delta_n(x^n,y^n) =\frac{1}{n}
\sum_{i=1}^n \Delta(x_i,y_i)$ and let  $R_{F,n}(D)$ and $ R_{O,n}(D)$
be the $n$th-order functional and operational rate-distortion
functions, respectively, defined by replacing  $\Delta$ with
$\Delta_n$,  $X$ with $X^n$,  $Y$ with $Y^n$ in \eqref{eq:def:frdf}
and \eqref{eq:onerddef},  and 
normalizing by the block-length $n$. Then
Theorem~\ref{thm:one_shot_rd_fn} gives the following corollary. 
\begin{corollary} \label{cor:one_shot_rd_fn}
  For any block distribution $P_{X^n}$ and any $D > 0$,
\begin{equation}
R_{F, n}(D) \le R_{O, n}(D) \le R_{F, n}(D) + \frac{\log e + 1}{n}.
\end{equation}
\end{corollary}

\begin{remark} \label{rem:entropy-code} Note that for simplicity  we have used the
  Shannon entropy $H(Y)$ as the rate of the code when defining $R_O(D)$. If,
  instead, one defined the rate as the expected codeword length  $\Exp[\length{\enc(Y)}]$ of a uniquely decodable description $\enc(Y)$, then the
  constant $1$ in the upper bounds of Theorem~\ref{thm:one_shot_rd_fn}  and
  Corollary~\ref{cor:one_shot_rd_fn} would become $2$.  For 
  example, in this case  the upper bound in  Theorem~\ref{thm:one_shot_rd_fn}
  would be $ R_O(D) \le R_F(D) + \log e + 2$. 
  
\end{remark}
\begin{proof}[Proof of Theorem~\ref{thm:one_shot_rd_fn}]
For the lower bound, every feasible countably valued $Y$ satisfies
$R_F(D)\le \FI{X}{Y}\le H(Y)$, where the second inequality follows from
Theorem~\ref{thm:fi_lb} and $H(Y\mid Z)\le H(Y)$ for any functional
representation. For the upper bound we use the following lemma from the
proof of~\cite[Thm.~2]{li2018strong}.
\begin{lemma}[Randomization between two quantizers] \label{lemma:derandom} For arbitrary
  source and reproduction alphabets $\XSpace$ and $\YSpace$ and
  distortion measure $\Delta : \XSpace \times \YSpace \to [0,\infty)$,
  let $Z$ be independent of $X$ and $Y = g(X, Z)$ for some measurable $g$.
  If $\Exp[\Delta(X, Y)] \le D$ and $H(Y \mid Z) < \infty$, then there
  exist measurable maps $g_0,g_1 : \XSpace \to \YSpace$ with countable
  range and a Bernoulli variable $B\perp X$ such that, with
  $\widetilde Y=g_B(X)$,
        \begin{equation}
            \Exp[\Delta(X,\widetilde Y)] \le D
        \end{equation}
        and
        \begin{equation}
            H(B,\widetilde Y)=H(B)+H(\widetilde Y\mid B)\le 1+H(Y\mid Z).
        \end{equation}
  The encoder draws the binary selector $B$ privately and includes it in
  the description of $\widetilde Y$.
\end{lemma}
Let $D>0$ and fix $\epsilon>0$. We may assume $R_F(D)<\infty$.
Choose $P_{Y\mid X}$ such that $\Exp[\Delta(X,Y)]\le D$ and
$\FI{X}{Y}\le R_F(D)+\epsilon$. Then Theorem~\ref{thm:rtc_for_pfr}
gives common randomness $Z$ independent of $X$ such that $Y=g(X,Z)$
and $H(Y\mid Z)\le \FI{X}{Y}+\log e$. Applying
Lemma~\ref{lemma:derandom} yields a feasible reproduction $\widetilde Y$ with
\begin{equation*}
    R_O(D)\le H(\widetilde Y)\le H(B,\widetilde Y)\le R_F(D)+\log e+1+\epsilon,
\end{equation*}
which, because $\epsilon$ was arbitrary, gives the result.
\end{proof}

If $P_X$ is atomless (for example, if $X$ has a density on $\Reals^n$),
a deterministic quantizer suffices for the same upper bound. Indeed, on
each cell $\{g_0=a,g_1=b\}$, assign $b$ to the fraction $\Prob(B=1)$ of
the cell's mass with the smallest values of $\Delta(x,b)-\Delta(x,a)$,
splitting ties by atomlessness, and assign $a$ elsewhere. This preserves
the output distribution of $g_B(X)$ and can only decrease its expected distortion.

Theorem~\ref{thm:one_shot_rd_fn} demonstrates the importance of $R_F(D)$ to rate-distortion theory, motivating a thorough study of its properties. As a small first step toward this, we show that $R_F(D)$ has considerable structure when the spaces $\XSpace, \YSpace$ are finite. 
\begin{proposition}[Piecewise linearity of $R_F$]
\label{prop:frdf_piecewise_lin}
Let $\XSpace, \YSpace$ be finite.
Then, $R_F(D)$ is a  piecewise linear function of $D$ on $[0,\infty)$.
\end{proposition}
\begin{proof}
See Appendix~\ref{app:helper_proofs}.
\end{proof}
\subsection*{Examples of Functional Rate-Distortion Functions}
\textbf{Bernoulli Source with Hamming Distortion.} 
    Let $X$ be Bernoulli with parameter $0<p<1$, $X \sim \Bern(p)$, and consider the functional rate-distortion function with Hamming distortion constraint $\Prob(X \ne Y) \le D$. Then, it follows after computation that
    \begin{equation}
        R_F(D) = \begin{cases}
            h_b(p)\left(1 - \frac{D}{p_{\text{min}}} \right), &0 \le D < p_{\text{min}} \\
            0, &D \ge p_{\text{min}},
        \end{cases}
    \end{equation}
    where $p_{\text{min}} \defeq \min(p, 1-p)$ and $h_b(\cdot)$ is the binary entropy function. Comparing this to the classical rate-distortion function
    \begin{equation}
        R(D) = \begin{cases}
            h_b(p) - h_b(D), &0 \le D < p_{\text{min}} \\
            0, &D \ge p_{\text{min}},
        \end{cases}
    \end{equation}
    we see that 
    \begin{equation}
        \begin{cases}
            R_F(D) > R(D), &0 < D < p_{\text{min}} \\
            R_F(D) = R(D), &D = 0 \text{ or } D \ge p_{\text{min}}.
        \end{cases}
    \end{equation}
\textbf{$N$-ary Uniform Source.} Let $X$ be an $N$-ary uniform source with alphabet $\XSpace = \qty{1, 2, \ldots, N}$. Let $\YSpace = \XSpace \cup \qty{e}$, where $e$ is the erasure symbol, and define the distortion measure 
\begin{equation}
    \Delta(x, y) = \begin{cases}
        0, &x = y \\
        1, &y = e, \\
        \infty,&  y \in \XSpace,\,  y \ne x.
    \end{cases}
\end{equation}
For $0\le D\le1$, it is well known (e.g.,~\cite[Prob. 6.5]{alajaji2018introduction}) that the rate-distortion function is given by $R(D) = (1 - D) \log N$. In our case, the minimizing channel $P_{Y \mid X}$ is singular, and hence
\begin{equation}
    R_F(D) = R(D) = (1 - D) \log N.
\end{equation}
Moreover, this channel satisfies the convex hull condition of
Proposition~\ref{prop:perf_fun_rep}. The encoder can erase $X$ with
probability $D$, independently of $X$, and reproduce $X$ otherwise.
The resulting output has entropy $(1-D)\log N+h_b(D)$, so
\begin{equation}
    (1-D)\log N\le R_O(D)\le (1-D)\log N+1.
\end{equation}
For $D\ge1$, all three rate-distortion functions are zero. \par
We also remark that our rate-distortion framework immediately extends to the problem of approximate variable-length channel simulation. Setting the distortion to be, for example, the total variation distance with the desired conditional distribution, Theorem~\ref{thm:one_shot_rd_fn} shows that the optimal method for approximate channel simulation is to exactly simulate, using the Poisson functional representation, a `nearby' channel. 
\section{R\'enyi Generalization} \label{sec:generalizations}
As  in~\cite{hill2025communication, wei2026r}, one may be concerned about channel simulation under costs other than the expected message length. In particular, we study channel simulation under the Campbell cost~\cite{campbell1965coding}, an exponential storage cost appropriate for applications where buffer overflow can occur~\cite{jelinek1968buffer, 1056322,blumer1988renyi, somazzi2024nonlinear}, and therefore, the cost of long codewords is especially significant~\cite{courtade2014cumulant, courtade2014variable, saito2022non}. This setup motivates the definition of a R\'enyi functional information, which shares similar properties with functional information, including closely characterizing the minimum Campbell cost of any stochastic code.  
\begin{definition}[R\'enyi Channel Simulation Divergence]
    For two probability measures $P \ll Q$ with width function $w$ and $\alpha \in (0, 1)$, define the \textit{R\'enyi channel simulation divergence} $\RCSD{\alpha}{P}{Q}$ as
    \begin{equation}
        \RCSD{\alpha}{P}{Q} \defeq h_\alpha(w) = \frac{1}{1-\alpha} \log \left( \int_0^\infty w(\ell)^\alpha \, d\ell \right),
    \end{equation}
    where $h_\alpha$ is the differential R\'enyi entropy.
\end{definition}
\begin{definition}[R\'enyi Functional Information]
    For $\alpha \in (0, 1)$ and $X, Y \sim P_{X, Y}$, the \textit{R\'enyi functional information} is defined as
    \begin{equation}
        \RFI{\alpha}{X}{Y} \defeq \frac{1}{1-\alpha} \log \left( \Exp_Y \left[ 2^{(1-\alpha) \RCSD{\alpha}{P_{X \mid Y}}{P_X}} \right] \right).
      \end{equation}

More explicitly,
\[
\RFI{\alpha}{X}{Y} = \frac{1}{1-\alpha}\log\left(\int_{\YSpace} \int_0^{\infty} w_y(\ell)^\alpha \, d\ell \, dP_Y(y)\right),
\]
where  $w_y(\ell) = P_X \left( \RND{P_{Y \mid X}}{P_Y}(y \mid X) \ge
  \ell \right)$. 
\end{definition}
The R\'enyi functional information satisfies the basic properties that
one expects of a mutual information. Before presenting these
properties, we recall the definitions of the R\'enyi divergence and
the Sibson $\alpha$-mutual information~\cite{esposito2025sibson}. 
\begin{definition}[R\'enyi divergence]
    For $P \ll Q$ probability distributions and $\alpha \in (0,1)\cup(1,\infty)$, the R\'enyi divergence of order $\alpha$ between $P$ and $Q$ is defined as
    \begin{equation}
    \alphaD{\alpha}{P}{Q} \defeq
        \frac{1}{\alpha - 1} \log\left(\mathbb{E}_{X \sim Q} \left[\left(\RND{P}{Q}(X) \right)^\alpha\right]\right). 
\end{equation}
\end{definition}
\begin{definition}[Sibson $\alpha$-Mutual Information]
    For random variables $X, Y \sim P_{X,Y}$ and for any $\alpha \in (0,1)\cup(1,\infty)$, the Sibson $\alpha$-mutual information is defined as
 \begin{equation}
     \SMI{\alpha}{X}{Y} \defeq \inf_{Q_Y:\,P_Y\ll Q_Y} \, \alphaD{\alpha}{P_{X, Y}}{P_X \otimes Q_Y}. \label{eq:sibsondefinition}
 \end{equation}

\end{definition}
\begin{proposition}[Properties of the R\'enyi functional information] \label{prop:rfi_props}
    Let $X, Y \sim P_{X, Y}$ and $\alpha \in (0, 1)$. Then,
    \begin{itemize}
        \item \textbf{(Nonnegativity)} $\RFI{\alpha}{X}{Y} \ge 0$, with equality iff $X \perp Y$.
        \item \textbf{(Recovers $I_F$ as $\alpha \to 1$)} If $\RFI{\alpha_0}{X}{Y}<\infty$ for some $\alpha_0\in(0,1)$, then $\displaystyle{\lim_{\alpha \to 1} \RFI{\alpha}{X}{Y} = \FI{X}{Y}}$.
        \item \textbf{(Self-information)} For $X$ discrete, $\RFI{\alpha}{X}{X} = H_\alpha(X)$, the R\'enyi entropy of $X$.
    \end{itemize}
\end{proposition}
\begin{proof} See Appendix~\ref{app:renyi_proofs}.
\end{proof}
The R\'enyi functional information is connected to the Sibson $\alpha$-mutual information analogously to how functional information relates to the mutual information. 
\begin{proposition}[Relationship to the Sibson $\alpha$-Mutual Information] \label{prop:rfi_to_sibson}
  For any $\alpha \in (0, 1)$ and $X, Y \sim P_{X, Y}$ such that
  $\RFI{\alpha}{X}{Y}<\infty$,   we have 
        \begin{equation}
            \SMI{\frac{1}{\alpha}}{X}{Y} \le \RFI{\alpha}{X}{Y}.  \label{eq:sibsonsandwich}
        \end{equation}
        Equality holds if and only if  $X \to Y$ is singular.
\end{proposition}
\begin{proof}
     See Appendix~\ref{app:renyi_proofs}.
\end{proof}
The R\'enyi functional information has an operational characterization connected to the minimum Campbell (exponential) cost of channel simulation~\cite{campbell1965coding, hill2025communication}. 
\begin{definition}[Campbell Cost]
    Let $t > 0$ and let $M$ be a binary message. The \textit{Campbell cost} $L_t$ of order $t$ is defined as
    \begin{equation}
        L_t \defeq \frac{1}{t} \log \left( \Exp \left[2^{t \length{M}} \right] \right). \label{eq:campbelldef}
    \end{equation}
\end{definition}
As $t \to  0$ in~\eqref{eq:campbelldef}, we recover the expected
message length $\Exp[\length{M}]$ provided $L_{t_0}<\infty$ for some
$t_0>0$. Campbell~\cite{campbell1965coding} showed that for a random
variable $V$ encoded into a uniquely decodable binary message $M$,
$H_\alpha(V) \le L_t$ for $\alpha = \frac{1}{1 + t}$. A suitable
prefix code achieves $L_t<H_\alpha(V)+1$ whenever $H_\alpha(V)<\infty$. This provides a R\'enyi generalization of Shannon's lossless source coding theorem. \par
One can ask about the fundamental limits of relative entropy
coding/channel simulation under the Campbell cost, for which we have
the following result. Recall the definition of
a stochastic code  $(Z, \enc, \dec)$ for $P_{Y|X}$ in \eqref{eq:stoch-code}. 

The Campbell cost associated with such a stochastic code, as defined in~\eqref{eq:expcost}, is 
\begin{equation}
  \label{eq:expcost1}
L_t  = \frac{1}{t} \log \left( \Exp \left[2^{t
      \abs{\enc(X,Z) }} \right] \right),
  \end{equation}
  and the minimum Campbell cost over all stochastic
  encoders is (as seen in~\eqref{eq:optexpcost}) given by 
\begin{equation}
    \label{eq:optexpcost1}
   L^*_t(X\to Y) = \inf_{(Z,\enc,\dec)} L_t. 
 \end{equation}
The associated  conditional entropy  is the
Arimoto-R\'enyi  conditional entropy of order $\alpha \in (0,1)$~\cite{arimoto1977information}
\begin{equation}
        H_\alpha^{\mathcal{A}}(Y \mid Z) \defeq
        \frac{\alpha}{1-\alpha} \log \Exp_{Z} \left[ \left(\sum_{y}
            P_{Y \mid Z}(y \mid Z)^\alpha \right)^{\frac{1}{\alpha}}
        \right],
        \label{eq:arimotocre}
    \end{equation}
and we consider (as per~\eqref{eq:optARent}) 
\begin{align}
R^*_\alpha(X \to Y)= \inf_{(Z,\enc,\dec)} H_\alpha^{\mathcal{A}}(Y \mid Z)  = \inf_{(Z,f)}
  H_\alpha^{\mathcal{A}}(Y \mid Z),
\end{align}
where $(Z,f)$ ranges over the functional representations specified in Definition~\ref{def:funct-rep}.
It follows from \cite{campbell1965coding} (after accounting for the side information $Z$) that when
$\alpha =\frac{1}{t+1}$, these quantities are related as
\begin{equation}
\label{eq:minrenyicost}
  R^*_\alpha(X \to Y)\le  L^*_t(X\to Y) \le R^*_\alpha(X \to Y) +1.
\end{equation}
When $Z \perp Y$, $H_\alpha^{\mathcal{A}}(Y \mid Z)=H_\alpha(Y)$ and~\eqref{eq:minrenyicost} recovers the above R\'enyi generalization of Shannon's lossless coding theorem.
\par
The next theorem is the R\'enyi-Campbell generalization of our main result,
Theorem~\ref{thm:fi_rate_sandwich}. 

\thmrenyigen*
\begin{proof}
  See Appendix~\ref{app:renyi_proofs}. We note that the upper bound is
  achieved by the Poisson functional representation with any proposal
  $Q_Y$ such that $P_{Y \mid X = x} \ll Q_Y$ for $P_X$-a.e.~$x$.
\end{proof}
 
As $\alpha \to 1$ in~\eqref{eq:renyi_pfr_theorem}, the constant
tends to $\log e$. For the minimum Campbell cost $L^*_t(X\to Y)$,
the additive constant in the upper bound increases by one bit.

    In~\cite[Cor. 2]{hill2025communication} it was shown that for any $\alpha \in (0, 1)$ and $\epsilon > 0$, there exists an encoding of the Poisson functional representation satisfying
    \begin{equation}
        L_t \le (1 + \epsilon) \SMI{\frac{1 + \epsilon(1 - \alpha)}{\alpha}}{X}{Y} + c(\alpha, \epsilon),
    \end{equation}
    where $\alpha = \frac{1}{1 + t}$ and $c(\alpha, \epsilon)$ is a
    constant. Theorem~\ref{thm:renyi_pfr} provides a near-matching
    lower bound in terms of the R\'enyi functional information (and
    therefore, also the Sibson $(1/\alpha)$-mutual
    information). Theorem~\ref{thm:renyi_pfr} improves
    on~\cite{hill2025communication} by constructing tight upper and
    lower bounds. Recently,~\cite{fan2026exact} independently defined
    the R\'enyi channel simulation divergence and proved upper and
    lower bounds in the problem of exact sampling.  The upper bound is
    achieved using a layered sampling scheme with Poisson
    processes. However, this lower bound does not extend to all
    relative entropy codes like that of Theorem~\ref{thm:renyi_pfr},
    and it is not immediately clear if the (more complex) layered
    sampling scheme is constant-gap optimal in our setting. In particular, one
    would need to construct an encoding of the index $K$ which does
    not depend on the input $X$, which is unobservable by the decoder.
\section{Conclusion and Open Questions}
In this paper, we have introduced the functional information
$\FI{X}{Y}$ and shown that it characterizes the minimum one-shot
conditional entropy cost of exact channel simulation to within
$\log e$ bits. A coding distribution adapted to the common randomness
gives this upper bound for the Poisson functional representation
under any dominating proposal distribution, and rejection sampling
satisfies the same bound when the corresponding density ratios are
uniformly bounded. These results close the logarithmic gap in earlier
characterizations based on mutual information. We have also
characterized perfect functional representations for finite alphabets
and given a construction attaining the lower bound for every
binary-output channel. For lossy compression with private encoder randomness, we have obtained
bounds on the operational entropy rate, placing it within
$(\log e+1)/n$ bits per symbol of the functional rate-distortion
function at blocklength~$n$. Finally, our R\'enyi generalization
characterizes the minimum Campbell cost of channel simulation within
an additive constant depending only on the R\'enyi order.

\par
We believe the functional information and ring toss code have many
applications in one-shot information theory, ranging from source
coding to learned compression. Having established the importance of
the functional information and channel simulation divergence to
information theory, we would like to understand their basic
properties. It is also important to practical applications to develop
fast estimators of these measures, given that they often do not admit
simple formulas.

\section{AI Use Statement} 
AI tools (ChatGPT~5.5, 5.6, and~6) were used to simplify and correct
proofs. In particular, GPT~5.5 was used to isolate the single
inequality condition \eqref{eq:key_inequality_bathtub} in
\Cref{thm:fi_lb}, simplify the form of \Cref{prop:perf_fun_rep} and construct the example in \Cref{prop:no_perfect_fi}, and
GPT 5.6 and 6 (Astra) were used to correct and simplify the proof of
the key upper bound \Cref{thm:rtc_for_pfr},  originally written by the
first two coauthors. All derivations and final proofs were manually verified by the authors.
%
%


\appendices

\section{Additional Proofs} \label{app:helper_proofs}

\begin{lemma}[Reverse Bathtub Identity] \label{lemma:bathtub}
        Let $(\Omega, \mathcal{F}, \mu)$ be a probability space and let $f : \Omega \to [0, \infty)$ be integrable. Define the generalized inverse 
        \begin{equation}
            f^{-1}(u) \defeq \sup \qty{ t \ge 0 : \mu (\qty{\omega : f(\omega) \ge t}) \ge u}, 
        \end{equation}
        for $0 \le u \le~1$.
        Then, for every $0 \le p \le 1$,
        \begin{equation}
        \sup_{\substack{0 \le b \le 1 \\ \int b d\mu = p}} \int b f \, d\mu = \int_0^p f^{-1}(u) \, du.
        \end{equation}
    \end{lemma}
    \begin{proof}
        This is a standard result; see, e.g.,~\cite[Theorem 1.14]{lieb2001analysis} for the form of the maximizer. We include the proof for completeness. Let $b : \Omega \to [0, 1]$ be admissible. By the layer cake formula,
        \begin{align*}
            \int bf \, d \mu &= \int_0^\infty \int b(x) \Ind{f(x) \ge t} \, d\mu(x) \, dt \\
            &\le \int_0^\infty \min \qty{p, \mu (f \ge t)} \, dt,
        \end{align*}
        because $0 \le b \le 1$ and $0 \le \Ind{f \ge t}$ imply that
        \begin{equation*}
            \int b \Ind{f \ge t} \, d\mu \le \int b \, d\mu = p \text{ and } \int b \Ind{f \ge t} \, d\mu \le \mu(f \ge t). 
        \end{equation*}
        Then, we can write
        \begin{align*}
            \int_0^\infty \min \qty{p, \mu(f \ge t)} \, dt &= \int_0^\infty \int_0^p \Ind{u \le \mu(f \ge t)} \, du \, dt \\
            &= \int_0^p \int_0^\infty \Ind{t \le f^{-1}(u)} \, dt \, du \\
            &= \int_0^p f^{-1}(u) \, du. 
        \end{align*}
        We have shown that every admissible $b$ satisfies 
        \begin{equation*}
            \int bf \, d\mu \le \int_0^p f^{-1}(u) \, du.
        \end{equation*}
        All that remains is to show that this upper bound is attained. The case $p = 0$ is trivially true, so assume $0 < p \le 1$ and set
        \begin{equation*}
            a = f^{-1}(p), \quad \quad  A_+ = \qty{f > a}, \quad \quad A_0 = \qty{f = a}. 
        \end{equation*}
        Then,
        \begin{equation*}
            \mu(A_+) \le p \le \mu(A_+) + \mu(A_0).
        \end{equation*}
Here the upper bound follows from $\mu(f \ge a) \ge p$.  For
the lower bound, if $\mu(f>a)>p$, continuity from below would
imply $\mu(f \ge a+\epsilon)>p$ for some $\epsilon>0$,
contradicting the definition of $a$.

        Choose $\alpha \in [0, 1]$ so that
        \begin{equation*}
            \mu(A_+) + \alpha \mu(A_0) = p,
        \end{equation*}
        and define $b_p = \Ind{A_+} + \alpha \Ind{A_0}$. Then, $0 \le b_p \le 1$ and $\int b_p \, d\mu = p$ by construction. We compute
        \begin{align*}
            \int_0^p f^{-1}(u) \,du &= \int_0^\infty \min \qty{p, \mu(f \ge t)} \, dt \\
            &= ap + \int_a^\infty \mu(f \ge t) \, dt \tag{$\mu(f \ge t) < p$ for $t > a$} \\
            &= ap + \int (f - a) \Ind{A_+} \, d\mu \tag{layer-cake formula} \\
            &= a(p - \mu(A_+)) + \int_{A_+} f \, d\mu \\
            &= a(\alpha \mu(A_0)) + \int_{A_+} f \, d\mu \tag{by the choice of $\alpha$} \\
            &= \alpha \int_{A_0} f \, d\mu + \int_{A_+} f \, d\mu \tag{$f = a$ on $A_0$} \\
            &= \int b_p f \, d\mu.
        \end{align*}
        Thus, the upper bound is attained by $b_p$. 
    \end{proof}
\begin{proof}[Proof of Proposition~\ref{prop:csd_convexity}]
    We show, more generally, that the channel simulation divergence is separately convex in $P$ and $Q$ on the domain $P\ll Q$. For $w(\ell) = Q \left( \RND{P}{Q} \ge \ell \right)$ and $w^{-1}$ its generalized inverse, we recall the representation of Lemma~\ref{lemma:bathtub}
    \begin{equation}
        \int_0^p w^{-1}(u) \, du = \sup_{\substack{0 \le b \le 1 \\
            \int b \, dQ = p}} \int b(x) \, dP(x). \label{eq:bathtub_pt2}
    \end{equation}
    For $p\in[0,1]$, we define $T(p) \defeq \sup \qty{\int b \, dP : 0 \le b \le 1,
      \int b \, dQ = p}$, writing~\eqref{eq:bathtub_pt2} as
    \begin{equation}
        \int_0^p w^{-1}(u) \, du = T(p).
    \end{equation}
    Since $\int_0^1 w^{-1}(u)\,du=1$, $T$ is continuous and nondecreasing on $[0,1]$, with $T(0)=0$ and $T(1)=1$.
    For fixed $Q$, $T(p)$ is a supremum of linear functions of $P$ over a constraint set that does not involve $P$. Therefore, $T(p)$ is convex in $P$ for a fixed $Q$. Using the representation 
    \begin{equation}
        \CSD{P}{Q} \cdot \ln 2 = -1 + \int_0^1 \frac{T(p)}{p} \, dp 
    \end{equation}
    recalled in~\eqref{eq:csd_representation}
    and~\cite[Lem. 3.1.2, item 2]{flamich2024data}, we conclude that
    $\CSD{\, \cdot}{Q}$ is convex in $P$, since integration preserves convexity. \par
    For a fixed $P$ and $a\in[0,1]$, define the generalized inverse $T^{-1}(a) = \inf \qty{p\in[0,1] : T(p) \ge a}$.
    By continuity and monotonicity of $T$, $T^{-1}(a)\le p$ if and only if $a\le T(p)$. Hence,
    \begin{align}
        \CSD{P}{Q} \cdot \ln 2 + 1 &=\int_0^1 \frac{T(p)}{p} \, dp \nonumber \\
        &=  \int_{0}^1 \frac{1}{p} \int_0^1 \Ind{T^{-1}(a) \le p} \, da\, dp \nonumber \\
        &=\int_0^1 \int_{T^{-1}(a)}^1 \frac{1}{p} \, dp \, da \tag{Tonelli} \\
        &= \int_0^1 - \ln T^{-1}(a) \, da. \nonumber
    \end{align}
Note that we may write 
    \begin{equation}
        T^{-1}(a) = \inf \qty{\int b \, dQ : 0 \le b \le 1, \int b \, dP \ge a}.
    \end{equation}
    Indeed, any admissible $b$ satisfies $\int b\,dP\le T(\int b\,dQ)$, so $\int b\,dQ\ge T^{-1}(a)$.
    Conversely, the maximizer in Lemma~\ref{lemma:bathtub} at $p=T^{-1}(a)$ satisfies
    $\int b\,dQ=T^{-1}(a)$ and $\int b\,dP=T(T^{-1}(a))\ge a$.
    For each fixed $a\in(0,1]$, and a fixed $b$, $\int b \, dQ$ is linear in
    $Q$. As the constraint set does not depend on $Q$, $T^{-1}(a)$ is
    therefore concave in $Q$. As $- \ln $ is convex and decreasing, $-
    \ln T^{-1}(a)$ is convex in $Q$; integrating over $a$ preserves
    convexity and hence, $\CSD{P}{\cdot\, }$ is convex in $Q$.
  \end{proof}

  \begin{proof}[Proof of Proposition~\ref{prop:fi_relation_to_mi}]
Recall that $\FI{X}{Y} = \CSD{P_{X \mid Y}}{P_X \mid P_Y}$.
Then, from the lower bound in \Cref{prop:csd_kl_sandwich_bound} we immediately get
\begin{align*}
\FI{X}{Y} &= \CSD{P_{X \mid Y}}{P_X \mid P_Y} \\
&\geq \KLD{P_{X \mid Y}}{P_X \mid P_Y} \\
&= \MI{X}{Y}.
\end{align*}
Next, from the upper bound in \Cref{prop:csd_kl_sandwich_bound}, we get
\begin{align*}
\FI{X}{Y} &= \CSD{P_{X \mid Y}}{P_X \mid P_Y} \\
&\leq \Exp_{Y}\big[\KLD{P_{X \mid Y}}{P_X}  \\
&\quad\quad\quad+ \log(\KLD{P_{X \mid Y}}{P_X} + 1) + 1\big]\\
&\leq \MI{X}{Y} + \log(\MI{X}{Y} + 1) + 1.
\tag{Jensen}
\end{align*}
For the equality condition in the lower bound, since $\MI{X}{Y}<\infty$,
equality holds if and only if the conditional lower bound is an equality
for $P_Y$-a.e.~$y$. Recalling that
\begin{align}
w_y(h) = P_X\left(\RND{P_{X \mid Y = y}}{P_X} \geq h\right),
\end{align}
the proof of \Cref{prop:csd_kl_sandwich_bound} in~\cite{goc2024channel}
shows that the conditional lower bound is an equality if and only if
\begin{align*}
\int_0^h w_y(\ell) \, d\ell \geq h w_y(h)
\end{align*}
holds with equality for $w_y(h)\,dh$-a.e.~$h$.
Since $w_y$ is nonnegative, nonincreasing, and integrates to one,
this holds if and only if, for some $g(y)\in[1,\infty)$,
\[
w_y(h)=\frac{1}{g(y)}\Ind{0<h\le g(y)}
\quad\text{for a.e. }h>0.
\]
Indeed, equality at $h$ forces $w_y(\ell)=w_y(h)$ for
a.e.~$\ell\in(0,h)$.
The displayed form of $w_y$ is equivalent to
$\RND{P_{X\mid Y=y}}{P_X}(x)\in\{0,g(y)\}$ for $P_X$-a.e.~$x$,
or, equivalently,
$\RND{P_{X\mid Y=y}}{P_X}(x)=g(y)$ for $P_{X\mid Y=y}$-a.e.~$x$.
Finally,
$\RND{P_{X\mid Y}}{P_X}(x\cond y)
=\RND{P_{Y\mid X}}{P_Y}(y\cond x)$
for $P_{X,Y}$-a.e.~$(x,y)$, so the equality condition is precisely
that $X\to Y$ is singular.
\end{proof}
\begin{proof}[Proof of Proposition~\ref{prop:fi_dpi}]  
By the Markov property, $P_{X\mid Z}=\Exp[P_{X\mid Y}\cond Z]$,
where the conditional expectation is understood setwise. Thus,
\begin{align*}
\FI{X}{Z} 
&= \CSD{P_{X \mid Z}}{P_X \mid P_Z} \\
&= \CSD{\Exp[P_{X \mid Y} \mid Z]}{P_X \mid P_Z} \tag{Markov structure}\\
&\leq \CSD{P_{X \mid Y}}{P_X \mid P_{Y, Z}} \tag{Proposition~\ref{prop:csd_convexity}}\\
&= \CSD{P_{X \mid Y}}{P_X \mid P_Y}\\
&= \FI{X}{Y}. \tag*{\qedhere}
\end{align*}
\end{proof}%

\begin{proof}[Proof of Proposition~\ref{prop:fi_convex_concave}]

\noindent
\textbf{Concavity in $P_X$.}
Let $p \in (0, 1)$, set $\bar{p} = 1 - p$ for convenience and let $\theta \sim \Bern(p)$.
The cases $p=0$ and $p=1$ are immediate.
Now, for two distributions $P_0, P_1$ over $\XSpace$, let $X \mid \theta \sim P_\theta$.
For $j \in \{0,1\}$, let $(X_j,Y_j)$ have source distribution $P_j$ and channel $P_{Y \mid X}$.
Now, note that $\theta \rightarrow X \rightarrow Y$ is a Markov chain. Therefore, 
\begin{align}
 \FI{\theta, X}{Y} &= \CSD{P_{\theta , X \mid Y}}{P_{\theta , X} \mid P_{Y}} \nonumber \\
&= \CSD{P_{\theta \mid X}{P_{X \mid Y}}}{P_{\theta \mid X}P_X \mid P_Y} \tag{$\theta \perp Y \mid X$} \\
&= \CSD{P_{X \mid Y}}{P_X \mid P_Y} \label{eq:width_fun_calculation} \\
&= \FI{X}{Y}, \nonumber
\end{align}
where~\eqref{eq:width_fun_calculation} holds because
\begin{align*}
    &\Prob_{\theta, X \sim P_{\theta, X}} \left( \RND{P_{\theta, X \mid Y}}{P_{\theta, X}}(\theta, X \mid y) \ge \ell \right) \\
    &= \Prob_{\theta, X \sim P_{\theta, X}} \left( \RND{P_{X \mid Y}}{P_{X}}(X \mid y) \ge \ell \right) \\
    &= w_y(\ell). 
\end{align*}
For $j\in\{0,1\}$, put $c_j(y)=\RND{P_{Y_j}}{P_Y}(y)$, which exists since
$P_Y=\bar p P_{Y_0}+pP_{Y_1}$. The chain rule gives, for $P_Y$-a.e.~$y$
and $P_{\theta,X}$-a.e.~$(\theta,x)$,
\begin{equation*}
\RND{P_{\theta, X \mid Y}}{P_{\theta, X}}(\theta, x \mid y)
=c_\theta(y)\RND{P_{X \mid Y, \theta}}{P_{X \mid \theta}}(x \mid y, \theta),
\end{equation*}
where the right-hand side is taken as zero when $c_\theta(y)=0$.
Thus, for $h>0$,
\begin{align*}
w_{y}(h)
&= P_{\theta, X}\left(\RND{P_{\theta, X \mid Y}}{P_{\theta, X}}(\, \cdot \mid y) \geq h\right) \\
&= P_{\theta, X}\left(c_\theta(y)\RND{P_{X \mid Y, \theta}}{P_{X \mid
    \theta}}(\, \cdot \mid y, \cdot\, ) \geq h\right) \\
&= \Exp_{\theta}[w_{y, \theta}(h/c_\theta(y))],
\end{align*}
where $w_{y,j}(h)=P_j\left(\RND{P_{X_j\mid Y_j=y}}{P_j}\geq h\right)$.
For $c_j(y)=0$, interpret $w_{y,j}(h/c_j(y))$ and
$c_j(y)\DiffEnt(w_{y,j})$ as zero. From this, via the differential entropy definition of the functional information, we get that
\begin{align*}
\FI{\theta, X}{Y}
  &=\Exp_{Y}[\DiffEnt(w_{Y})] \\*
&\ge \Exp_\theta\left[\Exp_Y[c_\theta(Y)\DiffEnt(w_{Y,\theta})]\right]
  \tag{concavity of $h(\,\cdot\,) $}  \\
&=\bar{p}\FI{X_0}{Y_0} + p\FI{X_1}{Y_1}.
\end{align*}
The last equality uses $c_j(y)\,dP_Y(y)=dP_{Y_j}(y)$.
Putting everything together, we have shown that
\begin{align*}
\FI{X}{Y}
&\geq \bar{p}\FI{X_0}{Y_0} + p\FI{X_1}{Y_1},
\end{align*}
where $P_X = \bar{p}P_0 + pP_1$, as desired.
\\
\noindent
\textbf{Convexity in $P_{Y \mid X}$.}
Let $p \in [0, 1]$, set $\bar{p} = 1 - p$ for convenience and let $\theta \sim \Bern(p)$ be independent of $X$.
Now, for two Markov kernels $P_0(x), P_1(x)$ over $\YSpace$, let $Y \mid X, \theta \sim P_\theta(X)$.
For $j \in \{0,1\}$, let $Y_j \mid X \sim P_j(X)$.
Since $P_{X\mid Y}=\Exp[P_{X\mid Y,\theta}\mid Y]$, we have
\begin{align*}
\bar{p}\FI{X}{Y_0} + p\FI{X}{Y_1}
&= \CSD{P_{X \mid Y, \theta}}{P_{X \mid \theta} \mid P_{Y, \theta}} \\
&= \CSD{P_{X \mid Y, \theta}}{P_{X} \mid P_{Y, \theta}} \tag{$X \perp \theta$} \\
&\geq \CSD{P_{X \mid Y}}{P_{X} \mid P_{Y}} \tag{Proposition~\ref{prop:csd_convexity}}\\
&=\FI{X}{Y}. \tag*{\qedhere}
\end{align*}
\end{proof}

\begin{proof}[Proof of Proposition~\ref{prop:frdf_piecewise_lin}]
    We use the following lemma.
    \begin{lemma}[Piecewise linearity and polyhedral convexity]
\label{lemma:fi_piecewise_linear}
Let $\XSpace = \{1, \ldots, n\}$ and $\YSpace = \{1, \ldots, m\}$, and fix $X \sim P_X$.
Identify $P_{Y \mid X}$ with its channel matrix $K = (K_{iy})$, where
$K_{iy} = P_{Y \mid X}(y \cond  i)$, in the polytope
\begin{align*}
\mathcal{K} \defeq \left\{K \in \Reals^{n \times m} : K_{iy} \geq 0,\;\;
    \sum_{y=1}^m K_{iy} = 1 , \;\;  i=1,\ldots n \right\}.
\end{align*}
Then $K \mapsto \FI{X}{Y}$ is continuous and piecewise linear on $\mathcal{K}$: there is a finite covering of $\mathcal{K}$ by closed polyhedral cells on each of which $\FI{X}{Y}$ is a linear function of $K$.
Its extension by $+\infty$ outside $\mathcal{K}$ is proper polyhedral convex, meaning that its epigraph is a polyhedron.
\end{lemma}
\begin{proof}[Proof of Lemma~\ref{lemma:fi_piecewise_linear}]
Put $p_i = P_X(i)$ and $\eta(s) = -s\log s$ for $s \in [0,1]$, with $0\log 0 = 0$, and define
\begin{align*}
q_y \defeq \sum_{i=1}^n p_i K_{iy}, \qquad
v_y(t) \defeq \sum_{i=1}^n p_i\Ind{K_{iy} \geq t}.
\end{align*}
For $q_y > 0$, the width function of $P_{X \mid Y=y}$ relative to $P_X$ is $w_y(\ell) = v_y(q_y\ell)$.
The definition of functional information and the substitution $t = q_y\ell$ therefore give
\begin{align}
\FI{X}{Y}
&= \sum_{y:q_y>0} q_y \int_0^\infty \eta(w_y(\ell))\,d\ell \nonumber\\
&= \sum_{y=1}^m \int_0^1 \eta(v_y(t))\,dt.
\label{eq:fi_finite_width}
\end{align}
Indeed, $v_y(t) = 0$ for $t > 1$; if $q_y = 0$, then $v_y(t) = 0$ for every $t > 0$, so that output contributes zero.

For each $y \in \YSpace$, choose a permutation $\pi_y$ of $\XSpace$ such that, writing $k_{y,i} \defeq K_{\pi_y(i),y}$,
\begin{align*}
k_{y,1} \geq \cdots \geq k_{y,n},
\end{align*}
and put
\begin{align*}
k_{y,n+1} = 0, \qquad S_{y,0} = 0, \qquad
S_{y,i} \defeq \sum_{j=1}^i p_{\pi_y(j)}.
\end{align*}
On each nonempty interval $k_{y,i+1} < t < k_{y,i}$, we have $v_y(t) = S_{y,i}$.
Consequently,
\begin{align}
\FI{X}{Y}
&= \sum_{y=1}^m \sum_{i=1}^n (k_{y,i} - k_{y,i+1})\eta(S_{y,i})
\label{eq:fi_sorted_channels}\\
&= \sum_{y=1}^m \sum_{i=1}^n K_{\pi_y(i),y}
    \bigl[\eta(S_{y,i}) - \eta(S_{y,i-1})\bigr].
\label{eq:fi_linear_channels}
\end{align}
For a fixed tuple $\Pi = (\pi_1, \ldots, \pi_m)$, the last expression is a linear function $L_\Pi(K)$, valid on the closed polyhedral cell
\begin{align*}
C_\Pi \defeq \left\{K \in \mathcal{K} : K_{\pi_y(1),y} \geq \cdots \geq K_{\pi_y(n),y}
    \text{ for every }y \in \YSpace\right\}.
\end{align*}
These at most $(n!)^m$ cells cover $\mathcal{K}$.
Ties cause no difficulty, since the corresponding intervals in~\eqref{eq:fi_sorted_channels} have zero length.
The linear formulas agree on overlapping cells, since each equals~\eqref{eq:fi_finite_width}.
The finite closed covering therefore also proves continuity on $\mathcal{K}$.

By Proposition~\ref{prop:fi_convex_concave}, the map $K \mapsto \FI{X}{Y}$ is convex on $\mathcal{K}$.
Its extension by $+\infty$ outside $\mathcal{K}$ is thus proper, convex, and piecewise affine on its effective domain.
By~\cite[Theorem~2.49]{rockafellar1998variational}, its epigraph is a polyhedron, proving the final assertion.
\end{proof}
Use the notation of Lemma~\ref{lemma:fi_piecewise_linear} and write
\begin{align*}
d(K) \defeq \sum_{i=1}^n \sum_{y=1}^m P_X(i)K_{iy}\Delta(i,y).
\end{align*}
For every feasible $D$, the set $\{K \in \mathcal{K} : d(K) \leq D\}$ is compact.
By Lemma~\ref{lemma:fi_piecewise_linear}, $\FI{X}{Y}$ is finite and continuous in $K$, so the infimum defining $R_F(D)$ is attained.
As usual, set $R_F(D) = +\infty$ when the constraint is infeasible.
Feasible distortion levels exist because $\Delta$ is finite on the finite alphabets.

By the same lemma, the epigraph
\begin{align*}
\mathcal{E} \defeq \left\{(K,t) \in \mathcal{K} \times \Reals : \FI{X}{Y} \leq t\right\}
\end{align*}
is a polyhedron, where the joint distribution of $(X,Y)$ is determined by $P_X$ and $K$.
Since $d(K)$ is linear, the set
\begin{align*}
\mathcal{P} \defeq \left\{(K,D,t) : (K,t) \in \mathcal{E},\ d(K) \leq D\right\}
\end{align*}
is also a polyhedron, being the intersection of a polyhedral cylinder and a closed half-space.
Attainment of the minimum gives
\begin{align}
\operatorname{epi} R_F
= \left\{(D,t) : \text{there exists }K\text{ with }(K,D,t) \in \mathcal{P}\right\}.
\label{eq:frdf_epigraph_projection}
\end{align}
A linear projection of a polyhedron is a polyhedron~\cite[Proposition~3.55(a)]{rockafellar1998variational}.
Thus $R_F$ is proper polyhedral convex, and~\cite[Theorem~2.49]{rockafellar1998variational} gives finitely many real coefficients $a_j,b_j$ such that
\begin{align*}
R_F(D) = \max_{1 \leq j \leq M}(a_jD+b_j)
\end{align*}
on its domain $[0,\infty)$. 
This proves piecewise affinity with finitely many pieces, which is the meaning of piecewise linearity in the proposition.
The displayed representation also proves continuity on this domain,
including right continuity at $0$; the nonincreasing property follows from nesting of the feasible sets.
\end{proof}

\section{Perfect Functional Representations} \label{app:perfect_functional}
In this appendix, we will provide proofs of the results in
Section~\ref{sec:perfect_functional}. We first explicitly state the
condition on a functional representation $(Z, f)$ for it to be
perfect.

\begin{lemma} \label{lemma:fun_rep_equality_condition}
    Let $X, Y \sim P_{X, Y}$ with $\MI{X}{Y}<\infty$, and let $Z \perp
    X$ be such that $Y = f(X, Z)$ for a measurable $f$ and $P_{Y\mid
      Z=z}$ is purely atomic for $P_Z$-a.e.~$z$. Let $\nu_y$ be a
    regular conditional distribution of $Z$ given $Y=y$, and put
    $q_y(z)=P_{Y\mid Z=z}(\{y\})$, as in the proof of
    Theorem~\ref{thm:fi_lb}. The representation $(Z, f)$ is perfect if
    and only if, for $P_Y$-a.e.~$y$, there exists a measurable family
    of bathtub maximizers $b_{y,p}:\XSpace\to[0,1]$, $p\in(0,1)$, for
    $r(y\mid\cdot\, )$ under the constraint $\int b_{y,p}\,dP_X=p$, as in Lemma~\ref{lemma:bathtub}, such that, for $dp\otimes\nu_y$-a.e.~$(p,z)$,
    \begin{equation}
        \int_{\XSpace} b_{y, p}(x) \Ind{f(x, z) = y} \, dP_X(x) = \min \qty{p, q_y(z)}. \label{eq:bathtub_equality}
    \end{equation}
  \end{lemma}
  
\begin{proof}
    This condition follows immediately from the inequality~\eqref{eq:supremum_over_b} in the proof of Theorem~\ref{thm:fi_lb}. 
  \end{proof}

\begin{proof}[Proof of Proposition~\ref{prop:perf_fun_rep}]
    Recall the standing assumptions $P_X(x)>0$ and $P_Y(y)>0$ on the finite alphabets.
    First, suppose that $y$-compatible orderings and the probability mass function $\alpha$ on the admissible maps $\varphi : \XSpace \to \YSpace$ are given satisfying
    \begin{equation}
        P_{Y \mid X}(y \mid x) = \sum_{\varphi : \varphi(x) = y} \alpha(\varphi).
    \end{equation}
    We will show that $P_{X, Y}$ admits a perfect functional representation. Define $Z \perp X$ to take values in the space of admissible maps, with $\Prob(Z = \varphi) = \alpha(\varphi)$. Set $f(x, \varphi) = \varphi(x)$. Then, $(Z, f)$ is a functional realization of the channel. Indeed, for each $x \in \XSpace, y \in \YSpace$,
    \begin{align*}
        \Prob(f(x, Z) = y) &= \sum_{z : f(x, z) = y} \Prob(Z = z) \\
        &= \sum_{\varphi : \varphi(x) = y} \alpha(\varphi) = P_{Y \mid X}(y \mid x).
    \end{align*}
    We next show that $(Z, f)$ satisfies the equality condition in Lemma~\ref{lemma:fun_rep_equality_condition}. Fix $y \in \YSpace$. For $0 \le p < 1$, let
 $        k = \max \big\{i\in\{0,\ldots,n-1\}\, :\, \sum_{j=1}^i
 P_X(x_{y,j})\le p\big\}$,     where the empty sum is zero. Define $A_{y, p} = \qty{x_{y, 1}, \ldots, x_{y, k}}$, with $A_{y,p}=\emptyset$ when $k=0$, and put
    \begin{equation}
        b_{y, p}(x) = \Ind{A_{y, p}}(x) + \frac{p - P_X(A_{y, p})}{P_X(x_{y, k+1})}\Ind{x=x_{y,k+1}},
    \end{equation}
and set $b_{y,1}(x)=1$ for all $x$. Then, $0 \le b_{y, p} \le 1$ and $\int_{\XSpace} b_{y, p}(x) \, dP_X(x) = p$. Because the ordering is $y$-compatible, this is a measurable family of bathtub maximizers for $r(y\mid x)=P_{Y\mid X}(y\mid x)/P_Y(y)$.
    Let $\varphi$ be an admissible mapping with $\alpha(\varphi) > 0$. Recall that $\varphi^{-1}(y)$ is an initial segment of the ordering of $x$. If $p \le q_y(\varphi)$, then $\{b_{y,p}>0\}\subseteq\varphi^{-1}(y)$ and hence
    \begin{equation*}
        \sum_{x \in \XSpace} b_{y, p}(x) \Ind{\varphi(x)=y} P_X(x) = \sum_{x \in \XSpace} b_{y, p}(x) P_X(x) = p.
    \end{equation*}
    Alternatively, if $p > q_y(\varphi)$, then $\varphi^{-1}(y)\subseteq\{b_{y,p}=1\}$ and hence
    \begin{align*}
        &\sum_{x \in \XSpace} b_{y, p}(x) \Ind{\varphi(x)=y} P_X(x) \\
        &= \sum_{x \in \XSpace} \Ind{\varphi(x)=y} P_X(x)
        = q_y(\varphi).
    \end{align*}
    Therefore,
    \begin{equation*}
        \int_{\XSpace} b_{y, p}(x) \Ind{\varphi(x) = y} \, dP_X(x) = \min \qty{p, q_y(\varphi)},
    \end{equation*}
    and $(Z, f)$ is a perfect representation. \par
    Conversely, suppose that a representation $(Z, f)$ is perfect. Let
    $\Phi = \qty{\varphi : \XSpace \to \YSpace}$ be the set of all
    functions from $\XSpace$ to $\YSpace$. For any functional
    representation $Y = f(X, Z)$ with $Z \perp X$, we have $f(\cdot,
    z) \in \Phi$ for any $z \in \ZSpace$, thus $f(\cdot, Z)$ is a
    $\Phi$-valued random variable. For $\varphi \in \Phi$, let
    $\alpha(\varphi) = \Prob(f(\cdot, Z) = \varphi)$. Then, we can
    introduce an equivalent discrete representation with $\ZSpace =
    \Phi$, $\Prob(Z = \varphi) = \alpha(\varphi)$, $X$ and $Z$
    independent and $f(x, \varphi) = \varphi(x)$. This preserves
    $H(Y\mid Z)$, since the conditional distribution of $Y$ given $Z=z$ depends only on the map $f(\cdot,z)$. Thus the new representation is also perfect. By definition, such a representation satisfies~\eqref{eq:varphi_to_satisfy}. All that remains is to show that the maps with $\alpha(\varphi) > 0$ are admissible for some fixed $y$-compatible orderings. In the remainder of the proof, consider only maps with positive $\alpha$-probability. Recall that $\varphi$ is admissible if, for every $y \in \YSpace$,
    \begin{equation*}
        \varphi^{-1}(y) = \qty{x_{y, 1}, \ldots, x_{y, k_y}}
    \end{equation*}
    for an index $k_y \in \qty{0, \ldots, n}$ and ordering $x_{y,1}, \ldots, x_{y,n}$ satisfying
    \begin{equation}
        P_{Y \mid X}(y \mid x_{y, i}) \ge P_{Y \mid X}(y \mid x_{y, j}),
    \end{equation}
    for all $i < j$. \par
    Fix $y \in \YSpace$. Then
    \begin{equation*}
        q_y(\varphi)=P_{Y\mid Z=\varphi}(y)=P_X(\varphi^{-1}(y)),
        \qquad
        \nu_y(\varphi)=\frac{\alpha(\varphi)q_y(\varphi)}{P_Y(y)}.
    \end{equation*}
    Choose a family $b_{y,p}$ as in Lemma~\ref{lemma:fun_rep_equality_condition}. Since there are only finitely many maps, there is a full-measure set of $p\in(0,1)$ on which~\eqref{eq:bathtub_equality} holds simultaneously for every $\varphi$ with $\nu_y(\varphi)>0$. All choices of $p$ below are made in this set.
    We claim that, for $\varphi,\psi$ with $q_y(\varphi),q_y(\psi)>0$, $q_y(\varphi)\le q_y(\psi)$ implies $\varphi^{-1}(y)\subseteq\psi^{-1}(y)$. Indeed, if $q_y(\varphi)<q_y(\psi)$, choose $q_y(\varphi)<p<q_y(\psi)$. Then~\eqref{eq:bathtub_equality} gives
    \begin{equation}
        \varphi^{-1}(y) \subseteq \qty{b_{y, p} = 1} \subseteq \qty{b_{y, p} > 0} \subseteq \psi^{-1}(y).
    \end{equation}
    Similarly, if $q_y(\varphi)=q_y(\psi)=q$, then for $p<q$,
    \eqref{eq:bathtub_equality} implies $\qty{b_{y,p}>0}\subseteq\varphi^{-1}(y)\cap\psi^{-1}(y)$. Since $\int b_{y,p}\,dP_X=p$, this implies
    \begin{equation}
        P_X\left(\varphi^{-1}(y)\cap\psi^{-1}(y)\right)\ge p.
    \end{equation}
    Letting $p\uparrow q$ through the full-measure set gives that the intersection has mass at least $q$. Since both cells have mass $q$ and $P_X(x)>0$ for every $x$, they are equal. Thus, the sets $\varphi^{-1}(y)$ form a nested chain ordered by their masses $q_y(\varphi)$. \par
    Next, for any $\varphi$ with $q_y(\varphi)>0$, we claim that for every $x\in\varphi^{-1}(y)$ and $x'\notin\varphi^{-1}(y)$,
    $P_{Y\mid X}(y\mid x)\ge P_{Y\mid X}(y\mid x')$. Suppose, toward a contradiction, that the reverse strict inequality holds for some such $x,x'$. Let $q=q_y(\varphi)$ and choose $p\in(q,q+P_X(x'))$ in the full-measure set. By~\eqref{eq:bathtub_equality}, $\varphi^{-1}(y)\subseteq\{b_{y,p}=1\}$.
    As $b_{y,p}$ maximizes the integral against $P_{Y\mid X}(y\mid x)/P_Y(y)$, we must also have $b_{y,p}(x')=1$: otherwise, transferring a small amount of $P_X$-mass of $b_{y,p}$ from $x$ to $x'$ would increase the objective while preserving $\int b_{y,p}\,dP_X=p$. Therefore,
    \begin{equation}
        \int_{\XSpace} b_{y, p}(x) \, dP_X(x) \ge q + P_X(x') > p,
    \end{equation}
    contradicting the definition of $b_{y,p}$. This establishes the claim. \par
    We now construct the $y$-compatible ordering of $\XSpace$. Sort $\XSpace$ according to decreasing $P_{Y\mid X}(y\mid x)$. Within each tie group, order the points by when they enter the nested chain of cells $\varphi^{-1}(y)$, placing points that never enter last and breaking any remaining ties arbitrarily. This ordering is, by definition, $y$-compatible. By the two claims, every $\varphi^{-1}(y)$ contains all points preceding any of its points, so it is an initial segment. \par
    Finally, if $\alpha(\varphi)>0$ but $\nu_y(\varphi)=0$, then $q_y(\varphi)=0$ and, since $P_X(x)>0$ for every $x$, $\varphi^{-1}(y)=\emptyset$, which is an initial segment of any ordering.
  \end{proof}%

\begin{proof}[Proof of Proposition~\ref{prop:bin_perf_rep}]
    Let $\YSpace = \qty{0, 1}$, $Z \sim \Unif(0, 1)$, and let $f(x, z) = \Ind{P_{Y \mid X}(1 \mid x) \ge z}$. Then, for every $x$,
    \begin{align*}
        \Prob(f(x, Z) = 1) &= P_{Y \mid X}(1 \mid x),
    \end{align*}
    and therefore $Y = f(X, Z)$. For $0 \le \ell \le 1/P_Y(y)$, we can then compute
    \begin{align*}
        w_y(\ell) &= P_X \left( \RND{P_{Y \mid X}}{P_Y}(y \mid X) \ge \ell \right) \\
        &= P_X \left( P_{Y \mid X} (y \mid X) \ge \ell P_Y(y) \right) \\
        &= \begin{cases}
            P_{Y \mid Z}(1 \mid \ell P_Y(1)), & y=1, \\
            P_{Y \mid Z}(0 \mid 1-\ell P_Y(0)), & y=0,
        \end{cases}
    \end{align*}
    where the last equality holds for a.e.~$\ell$. Substituting $z=\ell P_Y(1)$ for $y=1$ and $z=1-\ell P_Y(0)$ for $y=0$, we obtain that
    \begin{align*}
        &\FI{X}{Y} \\*
        &= -\sum_{y = 0}^1 P_Y(y) \int_0^{1 / P_Y(y)} w_y(\ell) \log w_y(\ell) \, d\ell \\
        &= -\sum_{y = 0}^1 P_Y(y) \int_0^1 P_{Y \mid Z}(y \mid z) \log P_{Y \mid Z}(y \mid z) \frac{1}{P_Y(y)} \, dz \\
        &= \Exp_Z \left[ h_b\left(P_{Y \mid Z}(1 \mid Z)\right) \right] \\
        &= H(Y \mid Z).
    \end{align*}
    We conclude that $(Z, f)$ is a perfect functional representation.
\end{proof}

\begin{proof}[Proof of Proposition~\ref{prop:no_perfect_fi}]
    Let $\YSpace = \XSpace = \qty{1, 2, 3}$ and define 
\begin{equation*}
P_{Y \mid X} = 
\begin{bmatrix}
1-\epsilon & 0 & \epsilon \\
\epsilon & 1-\epsilon & 0 \\
0 & \epsilon & 1-\epsilon
\end{bmatrix},
\end{equation*}
where $1/3 < \epsilon < 1/2$. We claim that no perfect functional representation exists for this channel. Let $(Z, f)$ be a functional representation of $Y$. As in the proof of Proposition~\ref{prop:perf_fun_rep}, we will replace this with an equivalent discrete representation $\ZSpace = \Phi$ with $f(x, \varphi) = \varphi(x)$ and $\Prob(Z = \varphi) = \alpha(\varphi)$. The $y$-compatible orderings are unique and are
\begin{equation}
    \begin{cases}
        1, 2, 3, &y = 1 \\
        2,3,1, &y = 2 \\
        3, 1, 2, &y = 3. 
    \end{cases} \label{eq:y_compatible_ordering}
\end{equation}
Since $P_{Y \mid X}(2 \mid 1) = P_{Y \mid X}(3 \mid 2) = P_{Y \mid X}(1 \mid 3) = 0$, all $\varphi$ with $\alpha(\varphi) > 0$ must satisfy
\begin{equation*}
    \varphi(1) \in \qty{1, 3}, \quad  \varphi(2) \in \qty{1, 2}, \quad \varphi(3) \in \qty{2, 3}.
\end{equation*}
Therefore, the only support-compatible maps are
\begin{align*}
    &(1, 1, 2), (1, 1, 3), (1, 2, 2), (1, 2, 3), \\
    &(3, 1, 2), (3, 1, 3), (3, 2, 2), (3, 2, 3),
\end{align*}
where a triple denotes $(\varphi(1), \varphi(2), \varphi(3))$. Among these, the only maps admissible for the $y$-compatible ordering in~\eqref{eq:y_compatible_ordering} are
\begin{equation*}
    \varphi_1 = (1, 1, 3), \varphi_2 = (1 , 2, 2), \varphi_3 = (1, 2,3), \varphi_4 = (3, 2, 3).
\end{equation*}
Indeed, their cells are
\[
\begin{array}{c|ccc}
  & \varphi^{-1}(1)&\varphi^{-1}(2)&\varphi^{-1}(3)\\ \hline
  \varphi_1  & \{1,2\} & \emptyset & \{3\}\\
  \varphi_2 & \{1\} & \{2,3\} & \emptyset\\
  \varphi_3 & \{1\} & \{2\} & \{3\}\\
  \varphi_4 & \emptyset & \{2\} & \{1,3\}
\end{array}.
\]
For $(Z, f)$ to be perfect, the convex hull requirement of Proposition~\ref{prop:perf_fun_rep} must hold for some weights $\alpha_1, \ldots, \alpha_4$. This means 
\begin{equation*}
    \alpha_4 = P_{Y \mid X}(3 \mid 1) = \epsilon,
\end{equation*}
as the only map which sends input $1$ to output $3$ is $\varphi_4$. Similarly, we must have that
\begin{equation*}
    \alpha_1 = P_{Y \mid X}(1 \mid 2) = \epsilon, \quad \alpha_2 = P_{Y \mid X}(2 \mid 3) = \epsilon.
\end{equation*}
Therefore,
\begin{equation*}
    \alpha_1 + \alpha_2 + \alpha_3 + \alpha_4 \ge 3 \epsilon > 1,
\end{equation*}
violating the requirement that $\alpha$ be a probability distribution. We conclude that a perfect functional representation cannot exist for this channel.
\end{proof}

\section{The Generalized Ring Toss Code for Rejection Sampling} \label{app:generalised_ring_toss}
Given a target distribution $P \ll Q$ with $r = \RND{P}{Q} \le M$, rejection sampling draws $\qty{(Y_i,U_i)}_{i \ge 1} \iid Q \otimes \Unif(0,1)$ and returns the index
\begin{equation*}
K = \min \qty{i : U_i \le \frac{1}{M} r(Y_i)}.
\end{equation*}
Then $Y_K \sim P$. For channel simulation, suppose $dP_{Y \mid X}/dP_Y \le M$ uniformly and use the shared randomness $Z = \qty{(Y_i,U_i)}_{i \ge 1} \iid P_Y \otimes \Unif(0,1)$. Given $X=x$, the encoder applies rejection sampling to $P_{Y \mid X=x}$ and sends a prefix-free description of $K$, allowing the decoder to recover the simulated output $Y=Y_K$.
As $K$ is geometrically distributed with success parameter $1/M$, which we write as $K \sim \Geom(1/M)$, we have
\begin{align}
\Ent(Y \mid Z) \leq \Ent(K \mid Z) \leq \Ent(K) \leq \log M + \log e.  
\end{align}
This ignores the decoder's knowledge of $Z$, and $\log M$ can greatly exceed $\FI{X}{Y}$. Conditioning the coding distribution on $Z$ gives the tighter bound below. Related sorting constructions achieve $\MI{X}{Y} + \Oh(\log (\MI{X}{Y} + 1))$ bits~\cite{phan2025channel,flamich2024data}.
\par
The conventional proposal $Q_Y=P_Y$ minimizes the conditional relative entropy:
\begin{equation*}
    \min_{Q_Y} \KLD{P_{Y \mid X}}{Q_Y \mid P_X} = \MI{X}{Y}.
\end{equation*}
Theorem~\ref{thm:rtc} instead gives a bound uniform over all admissible proposals, allowing $Q_Y$ to be chosen for ease of sampling or computational efficiency~\cite{flamich2023adaptive}.
\begin{theorem}[Constant-Gap Optimality of Rejection Sampling] 
\label{thm:rtc}
Given the joint distribution $P_{X, Y}$, let $Q_Y$ be any proposal distribution such that $P_{Y \mid X = x} \ll Q_Y$ for $P_X$-a.e.~$x$ and there exists $M > 0$ such that some version satisfies $\RND{P_{Y \mid X}}{Q_Y}(y \mid x) \le M$ for every $x, y$. 
Let $Z = \{(Y_i, U_i)\}_{i \ge 1} \iid Q_Y \otimes \Unif(0, 1)$ be independent of $X$ and define 
\begin{equation}
\label{eq:kdef}
K \defeq \min \left\{ i : M \,  U_i \le \RND{P_{Y \mid X}}{Q_Y}(Y_i \mid X) \right\}. 
\end{equation}
Then, $P_{Y_K \mid X=x}=P_{Y \mid X=x}$ for $P_X$-a.e.~$x$ and, with $Y=Y_K$,
\begin{equation}
    H(Y \mid Z) \le \FI{X}{Y} + \log e. \label{eq:rtc_codelength_ub}
\end{equation}
Moreover, this upper bound is achieved within one bit by encoding $K$
conditioned on $Z$ using the Shannon-Fano code  for the coding
distribution
\begin{equation}
Q_{K \mid Z}(k \mid z) \defeq w_{y_k}^Q(M u_k) \prod_{i=1}^{k-1} (1 - w_{y_i}^Q(M u_i)),
\end{equation}
for $w_y^Q$ a width function given by
\begin{equation}
w_y^Q(\ell) = P_X \left( \RND{P_{Y \mid X}}{Q_Y} (y \mid X) \ge \ell \right). \label{eq:qwidthfunction}
\end{equation} 
\end{theorem}
\begin{proof}
The correctness of this scheme, i.e., that $P_{Y_K \mid X=x}=P_{Y \mid X=x}$ for $P_X$-a.e.~$x$, follows from the correctness of standard rejection sampling. 
\par
Note that $Q_{K \mid Z}$ is a valid probability distribution, as $0 \le Q_{K \mid Z}(k \mid z) \le 1$ for each $k$ and, by using the identity that $w = 1 - (1 - w)$, we get
\begin{align*}
Q_{K \mid Z}(k \mid z) = \prod_{i=1}^{k - 1} (1 - w_{y_i}^Q(M u_i)) - \prod_{i=1}^{k} (1 - w_{y_i}^Q(M u_i)).
\end{align*}
This is a telescoping series, giving
\begin{align*}
\sum_{k=1}^\infty Q_{K \mid Z}(k \mid z) 
&= 1 - \lim_{n \to \infty}\prod_{i=1}^n (1 - w_{y_i}^Q(Mu_i)).
\end{align*}
However, note that the limit vanishes $P_Z$-a.e., as the $w_{Y_i}^Q(MU_i) \in [0, 1]$ are i.i.d. with mean $\frac{1}{M} > 0$. \par
Since $H(Y \mid Z) \le H(K \mid Z) \le H(P_{K \mid Z}; Q_{K \mid Z})$, we show~\eqref{eq:rtc_codelength_ub} by upper bounding the cross-entropy $H(P_{K \mid Z}; Q_{K \mid Z})$ between $Q_{K \mid Z}$ and the true conditional distribution $P_{K \mid Z}$ of $K$ given $Z$. We can compute
\begin{align}
H(P_{K \mid Z};Q_{K \mid Z})
&= \Exp_{K,Z}[- \log Q_{K \mid Z}(K \mid Z)] \nonumber \\
&= \Exp_{Y_K, U_K}[-\log w_{Y_K}^Q(MU_K) ]\label{eq:cross_entropy_terms} \\
&\quad- \Exp_{K,Z}\left[ \sum_{i=1}^{K-1} \log(1 - w_{Y_i}^Q(MU_i)) \right]. \nonumber
\end{align}
We will address each term in~\eqref{eq:cross_entropy_terms} separately; however, before doing so, let us determine the joint distribution of $(Y_K, U_K)$. We use the notation ${r^Q(y \mid x) \defeq \RND{P_{Y \mid X}}{Q_Y}}(y \mid x)$. By the definition of rejection sampling, we have
\begin{align}
\label{eq:rs_accept_coord_joint_distribution}
\frac{dP_{Y_K, U_K \mid X}}{d(Q_Y \otimes \Unif(0, 1))}(y, u \mid X) = M  \cdot \Ind{M u \leq r^Q(y \mid X)}.
\end{align}
Note that since we are using a universal upper bound $M$,  we have $K \mid X \sim \text{Geom}(1 / M)$ irrespective of $X$; i.e., $K$ conditioned on $X$ is geometrically distributed with parameter $1/M$. 
In particular, ${\Prob(K \geq k \mid X) = \left(1 - \frac{1}{M} \right)^{k-1}}$ and $\Exp[K \mid X] = M$.
In expectations over proposals below, $(Y, U) \sim Q_Y \otimes \Unif(0, 1)$ is independent of $X$. With $0\cdot(-\infty)=0$, we get
\begin{align}
&\Exp_{Y_K, U_K}[- \log\,w_{Y_K}^Q(M U_K)] \nonumber \\
&= \Exp_X\left[\Exp_{Y_K, U_K \mid X}[-\log w_{Y_K}^Q(MU_K) \mid X] \right] \nonumber \\
&= -\Exp_{Y, U, X}\left[M\Ind{MU \leq r^Q(Y \mid X)} \log w_{Y}^Q(MU)\right] \tag{by~\eqref{eq:rs_accept_coord_joint_distribution}} \\
&=\! -\!\Exp_{Y, U}\!\left[\!M\!\Exp_{X \mid Y, U}\!\left[\Ind{\!MU \!\leq \!r^Q\!(Y \!\mid \!X)}\! \mid \!Y, \!U\right] \! \log w_{Y}^Q(MU)\!\right] \nonumber \\
&= -\Exp_{Y, U}\left[M\cdot w_{Y}^Q(MU)\log w_{Y}^Q(MU)\right] \tag{def.\ of $w_y^Q(\ell)$}\\
&= -\Exp_{Y}\left[\int_0^1 M\cdot w_{Y}^Q(Mu)\log w_{Y}^Q(Mu) \, du\right] \nonumber \\
&= -\Exp_{Y}\left[\int_0^M  w_{Y}^Q(\ell)\log w_{Y}^Q(\ell) \, d\ell\right] \tag{letting $\ell = Mu$} \\
&= - \int_{\YSpace} \int_0^\infty w_y \left( \frac{\ell}{\RND{P_Y}{Q_Y}(y)} \right) \log w_y \left( \frac{\ell}{\RND{P_Y}{Q_Y}(y)} \right) \, d\ell \, dQ_Y(y) \tag{def. of $w_y(h) = P_X \left( \RND{P_{Y \mid X}}{P_Y}(y \mid X) \ge h \right)$} \\
&= - \int_{\YSpace} \int_0^\infty w_y(h) \log w_y(h) \, dh \, \RND{P_Y}{Q_Y}(y) \, dQ_Y(y) \tag{letting $h = \ell / \RND{P_Y}{Q_Y}(y)$} \\
&= \FI{X}{Y}. \label{eq:rtc_crossent_first_term}
\end{align}
For $Q_Y$-a.e.~$y$ with $\RND{P_Y}{Q_Y}(y)=0$, $w_y^Q(\ell)=0$ for $\ell>0$; the integrands in the change of variables are defined to be zero there.
We next address the second term in~\eqref{eq:cross_entropy_terms}. Let ${R_i = \log (1 - w_{Y_i}^Q(MU_i)) \cdot(1 - \Ind{MU_i \leq r^Q(Y_i \!\mid\! X)})}$ and write
\begin{align}
&\Exp_{K,Z}\left[\sum_{i=1}^{K-1} \log(1-w_{Y_i}^Q(MU_i))\right] \nonumber \\
&= \mathbb{E}_{X}\left[\mathbb{E}\left[\sum_{i=1}^K R_i\,  \Big |\,  X\right]\right] \nonumber \\
&= \mathbb{E}_X\Big[\mathbb{E}[K|X]\mathbb{E}_{Y,U|X}[(1 - \Ind{MU \le r^Q(Y|X)}) \nonumber \\
&\qquad\qquad\qquad \cdot\log(1-w_Y^Q(MU))]\Big] \label{eq:wald} \\
&= M\mathbb{E}_{X,Y,U}\left[(1 \!- \!\Ind{MU \!\le \!r^Q(Y|X)}) \log(1-w_Y^Q(MU))\right] \nonumber \\
&= M\mathbb{E}_{Y,U}\Big[\mathbb{E}_{X|Y,U}[1 - \mathbf{1}\{MU \le r^Q(Y|X)\} \mid Y, U] \nonumber \\
&\qquad\qquad\qquad \cdot \log(1-w_Y^Q(MU))\Big] \nonumber \\
&= M \Exp_{Y, U} \left[ (1 - w_{Y}^Q(MU)) \log (1 - w_{Y}^Q(MU)) \right] \label{eq:defofwidth} \\
&\geq M \Exp_{Y}\left[-\int_0^1 w_{Y}^Q(Mu) \log e \, du \right]  \label{eq:logineq}\\
&= - \int_{\YSpace} \int_0^M w_y^Q(\ell) \log e \, d\ell \, dQ_Y(y) \tag{letting $\ell = Mu$} \\
&= - \log e \int_{\YSpace} \RND{P_Y}{Q_Y} \, dQ_Y(y) \nonumber \\
&= - \log e, \nonumber
\end{align}
where in \eqref{eq:wald} we applied  Tonelli's theorem to 
$\sum_{i\ge1}(-R_i)\Ind{K\ge i}$ (given $X$, the $R_i$ are identically
distributed and $R_i$ is independent of $\{K\ge i\}$)  and the fact that 
$\sum_{i\ge1}\Prob(K\ge i\cond  X)=\Exp[K\cond 
X]$. Equation~\eqref{eq:defofwidth} follows from the definition of
$w_y^Q$.  Equation~\eqref{eq:logineq} follows using the inequality
$(1 - x) \log (1 - x) \geq -x \log e$.
Combining~\eqref{eq:cross_entropy_terms},~\eqref{eq:rtc_crossent_first_term},
and the above inequality completes the proof.
\end{proof}
\begin{figure}
    \centering
    \includegraphics[]{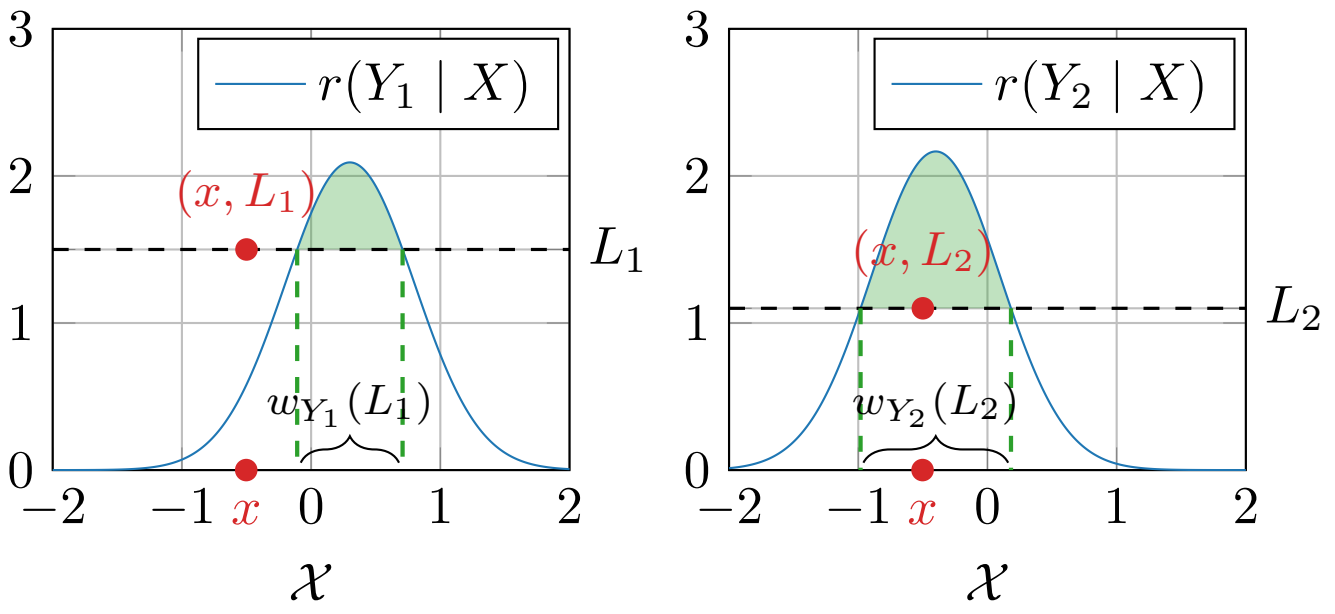}
    \caption{\textcolor{tabgreen}{Acceptance regions} for $Y=X+N$, with $X \sim \Normal{0}{0.75}$ and $N \sim \Normal{0}{0.25}$; $X$ and $Y$ are truncated to ensure a bounded density ratio. For \textcolor{tabred}{$x=-0.5$}, proposals $Y_1=0.3$, $Y_2=-0.4$ have heights $L_1=1.5$, $L_2=1.1$, where $L_i \defeq MU_i$. Since $x \notin S_1$ and $x \in S_2$, the first proposal is rejected and the second accepted. Each $S_i$ has $P_X$-probability $w_{Y_i}(L_i)$.}
    \label{fig:rts}
\end{figure}
Given $Z$, the decoder knows the acceptance sets $S_i = \qty{x \in \XSpace : \RND{P_{Y \mid X}}{Q_Y}(Y_i \mid x) \geq MU_i}$ illustrated in Fig.~\ref{fig:rts}. The acceptance test is
\begin{equation}
    M U_i \le \RND{P_{Y \mid X}}{Q_Y}(Y_i \mid x). \label{eq:acceptance_test_rs}
\end{equation}
Although the decoder does not know $x$, it can compute the marginal acceptance probability
\begin{equation}
    P_X \left( \RND{P_{Y \mid X}}{Q_Y}(Y_i \mid X) \ge M U_i \right) = w_{Y_i}^Q(M U_i).
\end{equation}
Since $K=k$ precisely when $x \in S_k$ and $x \notin S_i$ for $i<k$, the exact conditional distribution is
\begin{align}
P_{K \mid Z}(k \mid z) = P_X\left(S_k \setminus \bigcup_{i=1}^{k-1} S_i\right).
\end{align}
The coding distribution $Q_{K \mid Z}$ avoids computing these set differences by treating the accept/reject decisions as independent. The functional information lower bound implies that using the exact conditional distribution can reduce the cross-entropy by at most $\log e$ bits. \par
The uniform bound in Theorem~\ref{thm:rtc} allows the proposal to be chosen to minimize the expected number of proposals. With the smallest admissible rejection bound $M$, this expectation is
\begin{equation}
\label{eq:rtc_expected_runtime}
\Exp[K] = \esssup_{(x,y)\sim P_X\otimes Q_Y} \RND{P_{Y \mid X}}{Q_Y}(y \mid x).
\end{equation}
Minimizing~\eqref{eq:rtc_expected_runtime} over proposals dominating $P_{Y\mid X=x}$ for $P_X$-a.e.~$x$ gives the essential-supremum form of the Shtarkov solution~\cite{grunwald2009finiteness,shtar1987universal}:
\begin{equation}
\min_{Q_Y} \esssup_{(x,y)\sim P_X\otimes Q_Y} \RND{P_{Y \mid X}}{Q_Y}(y \mid x) = \Exp_{Y \sim P_Y} \left[\esssup_{x\sim P_X} \RND{P_{Y \mid X}}{P_Y}(Y \mid x)\right] \defeq M^*. 
\label{eq:shtarkov_minimization}
\end{equation}
When $M^*<\infty$, the unique minimizing proposal is the Shtarkov distribution $Q_Y^*$, defined for any $\sigma$-finite dominating measure $\nu$ with $P_{Y \mid X=x} \ll \nu$ $P_X$-a.s. by
\begin{equation}
\RND{Q_Y^*}{\nu}(y) = \frac{\esssup_{x\sim P_X} \RND{P_{Y \mid X}}{\nu}(y \mid x)}{\int \esssup_{x\sim P_X} \RND{P_{Y \mid X}}{\nu} (y \mid x) \, d \nu(y)}. \label{eq:shtarkov_distribution}
\end{equation}

Thus $M^*<\infty$ guarantees an admissible proposal with rejection bound $M^*$ and yields the following result.

\begin{proposition}[Shtarkov distribution is optimal for rejection sampling]
\label{thm:shtarkov_rej_samp}
Let $P_{X, Y}$ be such that the Shtarkov integral $M^*$ of $P_{Y \mid X}$, as given in~\eqref{eq:shtarkov_minimization}, is finite. Then, the expected number of proposals in rejection sampling is minimized by setting $Z = \qty{(Y_i, U_i)}_{i \ge 1} \iid Q_Y^* \otimes \Unif(0, 1)$, for $Q_Y^*$ defined in~\eqref{eq:shtarkov_distribution}. With $Z\perp X$ and rejection bound $M=M^*$, 
\begin{equation}
    \Exp[K] = \Exp_{Y \sim P_{Y}} \left[ \esssup_{x\sim P_X} \RND{P_{Y \mid X}}{P_Y}(Y \mid x) \right],
\end{equation}
and $H(Y \mid Z) \le \FI{X}{Y} + \log e$.
\end{proposition}
\section{Ring Toss code for Greedy Poisson Rejection Sampling} \label{app:gprs}

Let $X, Y \sim P_{X, Y}$ and consider 
\begin{align*}
r(y \mid x) &= \frac{dP_{Y \mid X}}{dP_Y}(y \mid x), \\
w_y(h) &= \Prob_{X \sim P_X}[r(y \mid X) \geq h], \\
w^x(h) &= \Prob_{Y \sim P_Y}[r(Y \mid x) \geq h]. 
\end{align*}
Let $G_x(h)=\int_h^\infty w^x(\eta)\,d\eta$.
Then, define
\begin{align*}
\sigma^x(h) &\defeq \int_0^h \frac{1}{G_x(\eta)} \, d\eta.
\end{align*}
We take $\sigma^x(h)=+\infty$ when $G_x(h)=0$. Let $Z = \{(Y_k, T_k)\}_{k = 1}^\infty$ be a Poisson process independent of $X$ with mean measure $P_Y \otimes \lambda$, indexed in increasing order of $T_k$, where $\lambda$ is Lebesgue measure on $[0,\infty)$.
Then, greedy Poisson rejection sampling (GPRS) returns~\cite{flamich2023greedy}
\begin{align*}
K(x, Z) \defeq \min\left\{k \in \Nats \mid T_k \leq \sigma^x(r(Y_k \mid x))\right\}.
\end{align*}
We write $K_x=K(x,Z)$ and $K=K(X,Z)$.
For $P_X$-a.e.~$x$, $Y_{K_x} \sim P_{Y \mid X = x}$ (see \cite{flamich2023greedy,flamich2024data}).
Write $Y=Y_K$ for the simulated output and $S_x(t)=\Prob[T_{K_x}\ge t]$. The joint density of the first accepted point is
\begin{align*}
\frac{dP_{Y_K, T_{K} \mid X}}{dP_Y \otimes \lambda}(y, t \mid x) 
&= \Ind{t \leq \sigma^x(r(y \mid x))} S_x(t), 
\end{align*}
where $S_x(t)=G_x((\sigma^x)^{-1}(t))$ by~\cite[Appendix~A, Eq.~(50)]{flamich2023greedy}.
From this, we can easily work out the conditional distribution of $T_K \mid Y_K, X$:
\begin{align*}
\frac{dP_{Y_K, T_{K} \mid X}}{dP_{Y} \otimes \lambda}(y, t \mid x) 
&= \frac{dP_{Y_K, T_{K} \mid X}}{dP_{Y_K \mid X} \otimes \lambda}(y, t \mid x) \cdot \frac{dP_{Y_K \mid X}}{dP_{Y}}(y \mid x)
\\
&=\frac{dP_{T_K \mid Y_K, X}}{d\lambda}(t \mid y, x) \cdot r(y \mid x).
\end{align*}
Hence, for $P_{X,Y}$-a.e.~$(x,y)$, where $r(y\mid x)>0$,
\begin{align*}
\frac{dP_{T_K \mid Y_K, X}}{d\lambda}(t \mid y, x) = \frac{\Ind{t \leq \sigma^x(r(y \mid x))}}{r(y \mid x)} S_x(t).
\end{align*}
For $t>0$, taking expectation  with respect to $P_{X|Y=y}$  gives
\begin{align*}
&\frac{dP_{T_K \mid Y}}{d\lambda}(t \mid y)\\
&= \Exp_{X \sim P_{X \mid Y = y}}\left[\frac{\Ind{t \leq \sigma^X(r(y \mid X))}}{r(y \mid X)} S_X(t) \right] \\
&= \Exp_{X \sim P_{X}}\left[\Ind{t \leq \sigma^X(r(y \mid X))} S_X(t) \right].
\end{align*}
For convenience, let $f(t \mid y) = \frac{dP_{T_K \mid Y}}{d\lambda}(t \mid y)$.
Now, put
\begin{align*}
\zeta(Z) &\defeq \sum_{(Y, T) \in Z} f(T \mid Y), \\
Q_{K \mid Z}(k) &\defeq \frac{f(T_k \mid Y_k)}{\zeta(Z)}.
\end{align*}
Since $T_K$ given $Y_K=y$ has PDF $f(\, \cdot\mid y)$, we have
$\zeta(Z)\ge f(T_K\cond  Y_K)>0$ a.s. Also, $\zeta(Z) < \infty$ a.s., because
\begin{align*}
\Exp[\zeta(Z)]&= \Exp\left[\sum_{(Y, T) \in Z} f(T \mid Y)\right] \\
&= \int\int_0^\infty f(t \mid y) \,dt \, dP_{Y}(y) \tag{Campbell}\\
&= 1. \tag{$f$ is a PDF in $t$}
\end{align*}
Thus, $Q_{K \mid Z}$ is a proper probability distribution. Furthermore,
\begin{align*}
\CrossEnt{P_{K \mid Z}}{Q_{K \mid Z}}
&= \Exp[-\log f(T_K \mid Y_K)] + \Exp[\log \zeta(Z)] \\
&\leq \Exp[-\log f(T_K \mid Y_K)] &\tag{Jensen and $\Exp[\zeta(Z)] = 1$} \\
&= \DiffEnt(T_K \mid Y_K).
\end{align*}
Put $H_x=(\sigma^x)^{-1}(T_{K_x})$. In the joint density of $(Y_K,T_K)$ given $X=x$, change variables from $t$ to $h$ via $t=\sigma^x(h)$. Since $S_x(\sigma^x(h))(\sigma^x)'(h)=1$ for $G_x(h)>0$, we obtain
\begin{align*}
P_{Y_K,H_X\mid X=x}(dy,dh)=\Ind{0<h\le r(y\mid x)}\,P_Y(dy)\,dh.
\end{align*}
Integrating out $y$ or $x$, respectively, shows that $H_X$ given $X=x$ has PDF $w^x$, and $H_X$ given $Y=y$ has PDF $w_y$. Since $G_x\le1$ is nonincreasing,
\begin{align*}
H_X \le T_K &= \sigma^X(H_X) \\
&\leq H_X \cdot (\sigma^X)'(H_X) \\
&=\frac{H_X}{G_X(H_X)}.
\end{align*}
By the probability integral transform, $G_X(H_X)$ given $X$ is uniform on $(0,1)$.
To construct a comparison density, let $U\sim\Unif(0,1)$ be independent of $(H_X,Y)$ and put $A=H_X/U$.
Let $g(a \mid y)$ denote the PDF of $A \mid Y = y$.
Then, one can find that
\begin{align*}
g(a \mid y)
&\defeq \frac{1}{a^2}\int_0^a h w_y(h) \, dh.
\end{align*}
At $y=Y_K$, we have
\begin{align*}
g(T_K \mid y) &= \frac{1}{T_{K}^2}\int_0^{T_K} h w_{y}(h) \, dh \\
&\geq \frac{1}{T_{K}^2}\int_0^{H_X} h w_y(h) \, dh \\
&\geq \frac{w_y(H_X)}{T_{K}^2}\int_0^{H_X} h \, dh \\
&= \frac{w_y(H_X)}{2} \cdot \left(\frac{H_X }{T_{K}}\right)^2 \\
&\geq  \frac{1}{2}w_y(H_X) G_X(H_X)^2.
\end{align*}
Finally, from this we get that
\begin{align*}
\DiffEnt(T_K \mid Y_K)
&\leq \Exp[-\log g(T_K \mid Y_K)] \\
&\leq \Exp[-\log w_{Y}(H_X)] - 2\Exp[\log G_X(H_X)] + 1\\
&= \DiffEnt(H_X \mid Y_K) + 2\log e + 1 \\
&= \FI{X}{Y} + 2\log e + 1.
\end{align*}
This shows that
\begin{align*}
\CrossEnt{P_{K \mid Z}}{Q_{K \mid Z}} \leq \FI{X}{Y} + 2\log e + 1.
\end{align*}
A Shannon-Fano code for $Q_{K\mid Z}$ therefore has expected length less than $\FI{X}{Y}+2\log e+2$.
\section{R\'enyi Generalization Proofs} \label{app:renyi_proofs}

\begin{proof}[Proof of Proposition~\ref{prop:rfi_props}] \mbox{\ }
 
\noindent \textbf{Nonnegativity.} By
Proposition~\ref{prop:rfi_to_sibson},
$\RFI{\alpha}{X}{Y}\ge\SMI{1/\alpha}{X}{Y}\ge0$, and the Sibson
information vanishes only when $X$ and $Y$ are
independent~\cite[Theorem~4.1]{esposito2025sibson}, in which case the
channel is singular, so equality holds throughout. \par
\noindent \textbf{Recovers $I_F$ as $\alpha \to 1$.} Let $V$ given $Y=y$ have density $w_y$, and put $B=-\log w_Y(V)$. Then $\RFI{1-s}{X}{Y}=s^{-1}\log\Exp[2^{sB}]$ for $0<s<1$. By hypothesis, $\Exp[2^{(1-\alpha_0)B}]<\infty$, so differentiating $\log\Exp[2^{sB}]$ at $s=0$ gives the limit $\Exp[B]=\FI{X}{Y}$. \par

\noindent \textbf{Self-information.} For $X$ discrete, fix $x$ with $P_X(x)>0$ and define $Q_x$ as the measure $Q_x(A) = \Ind{x \in A}$. One can compute that the width function $w_x$ between $Q_x$ and $P_X$ is given by 
    \begin{align*}
        w_x( \ell) &= \Prob_{X' \sim P_X} \left( \frac{\Ind{X' = x}}{P_X(X')} \geq \ell \right) \\
        &= P_X(x) \Ind{\ell \le \frac{1}{P_X(x)}}, \qquad \ell>0.
    \end{align*}
    Hence,
    \begin{equation*}
        \int_0^\infty w_x(\ell)^\alpha \, d\ell = \int_0^{\frac{1}{P_X(x)}} P_X(x)^\alpha \, d \ell = P_X(x)^{\alpha - 1}, 
    \end{equation*}
    and direct computation yields
    \begin{align}
        \RFI{\alpha}{X}{X} &= \frac{1}{1-\alpha} \log \Exp_X \left[ \int_0^\infty w_X(\ell)^\alpha \, d\ell \right] \nonumber \\
        &= \frac{1}{1-\alpha} \log \Exp_X \left[ P_X(X)^{\alpha-1} \,  \right] = H_\alpha(X). \tag*{\qedhere}
    \end{align}
  \end{proof}

\begin{proof}[Proof of Proposition~\ref{prop:rfi_to_sibson}]
    Put $p=1/\alpha>1$ and write $\norm{\, \cdot\, }_p$ for the $L^p(P_X)$ norm. By~\cite[Eq.~26]{esposito2025sibson},
    \begin{equation}
        \SMI{1/\alpha}{X}{Y} =
        \frac{1}{1-\alpha}\log\Exp_Y\!\left[\norm{r(Y\cond\cdot\, )}_p\right]. \label{eq:smiexpression}
    \end{equation}
    For $P_Y$-a.e.~$y$, the layer cake representation and Minkowski's integral inequality give
    \begin{align*}
        \norm{r(y\cond\cdot\, )}_p
        &= \left\|\int_0^\infty \Ind{r(y\cond\cdot\, )\ge\ell}\,d\ell\right\|_p \\
        &\le \int_0^\infty \norm{\Ind{r(y\cond\cdot\, )\ge\ell}}_p\,d\ell
        = \int_0^\infty w_y(\ell)^\alpha\,d\ell.
    \end{align*}
    Substitution into~\eqref{eq:smiexpression} gives~\eqref{eq:sibsonsandwich}.

    If the channel is singular, then for $P_Y$-a.e.~$y$ the nonzero
    level-set indicators agree $P_X$-a.s. for $\ell>0$, so equality
    holds in Minkowski's inequality  and hence
    in~\eqref{eq:sibsonsandwich}. Conversely, by finiteness, equality
    in~\eqref{eq:sibsonsandwich} requires equality in Minkowski's
    inequality  for $P_Y$-a.e.~$y$. For each such $y$, almost every
    level-set indicator with $w_y(\ell)>0$ is then proportional to
    $r(y\cond\cdot\, )$ in $L^p(P_X)$. Hence $r(y\cond\cdot\, )$ is
    equal to a  constant where it is positive, up to a $P_X$-null
    set. Since $\Exp_X[r(y\cond X)]=1$, this constant is
    $g(y)=1/\Prob_X[r(y\cond X)>0]$, which is measurable in $y$. This
    proves that the channel is singular. 
\end{proof}
\begin{proof}[Proof of Theorem~\ref{thm:renyi_pfr}]
    Let $Z \perp X$ be such that $Y = f(X, Z)$ and $P_{Y \mid Z = z}$ is purely atomic for $P_Z$-a.e.~$z$. We will show that $\RFI{\alpha}{X}{Y} \le H_\alpha^{\mathcal{A}}(Y \mid Z)$, where 
    \begin{equation*}
        H_\alpha^{\mathcal{A}}(Y \mid Z) =
        \frac{\alpha}{1-\alpha} \log \Exp_{Z} \left[ \left(\sum_{y}
            P_{Y \mid Z}(y \mid Z)^\alpha \right)^{\frac{1}{\alpha}}
        \right]
    \end{equation*}
    is Arimoto's conditional entropy~\cite{arimoto1977information}. Taking the infimum over these representations will give the lower bound. Following~\eqref{eq:key_inequality_bathtub} in the proof of Theorem~\ref{thm:fi_lb}, we may write
    \begin{equation}
        \int_0^p w_y^{-1}(u) \, du \le \int_{\ZSpace} \frac{\min \qty{p, q_y(z)}}{q_y(z)} \, d\nu_y(z), 
    \end{equation}
    for $q_y(z) = P_{Y \mid Z = z}(y)$ and $\nu_y$ a regular conditional distribution of $Z \mid Y = y$. For $P_Y$-a.e.~$y$, $0<q_y(z)\le1$ for $\nu_y$-a.e.~$z$. Extend $w_y^{-1}(u)=0$ for $u>1$. The inequality then also holds for $p\ge1$, since both sides equal one. Fixing such a $y$, we compute
    \begin{align*}
        \MoveEqLeft\int_0^\infty \alpha (1-\alpha) p^{\alpha - 2} \left(\int_0^p w_y^{-1}(u) \, du \right)\, dp \\
        &\leq \int_0^\infty \alpha(1-\alpha) p^{\alpha - 2} \left( \int_{\ZSpace} \frac{\min \qty{p ,q_y(z)}}{q_y(z)} \,d\nu_y(z) \right) \, dp\\
        &= \int_{\ZSpace} \int_0^\infty \alpha(1-\alpha) p^{\alpha - 2} \left(\frac{\min{\qty{p, q_y(z)}}}{q_y(z)} \right) \, dp \, d\nu_y(z) \tag{Tonelli}
        \\
        &= \int_{\ZSpace} q_y(z)^{\alpha - 1} \, d\nu_y(z).
    \end{align*}
    Above, the final equality follows from the identity (for any $q>0$ and $\alpha \in (0, 1)$)
    \begin{align*}
        \MoveEqLeft \alpha(1 - \alpha) \int_0^\infty p^{\alpha - 2} \min \qty{p, q} \, dp \\
        &= \alpha(1 - \alpha) \left[\int_0^q p^{\alpha - 1} \, dp  + \int_q^\infty q p^{\alpha - 2} \, dp \right] \\
        &= \alpha (1-\alpha) \left[ \frac{1}{\alpha} q^{\alpha} + \frac{1}{\alpha - 1} (-q^{\alpha - 1}) q \right] \\
        &= (1 - \alpha) q^{\alpha} + \alpha q^\alpha \\
        &= q^\alpha.
    \end{align*}
    Then, we can compute
    \begin{align*}
     \MoveEqLeft \alpha(1-\alpha) \int_0^\infty p^{\alpha-2} \left( \int_0^p w_y^{-1}(u) \, du \right) \, dp \\ &= \alpha(1-\alpha) \int_0^\infty w_y^{-1}(u) \int_u^\infty p^{\alpha -2} \, dp \, du \tag{Tonelli} \\
        &= \alpha(1-\alpha) \int_0^\infty w_y^{-1}(u) \left( \frac{u^{\alpha -1}}{1-\alpha} \right) \, du \\
        &= \int_0^\infty \int_0^{w_y(\ell)} \alpha u^{\alpha-1}\,du\,d\ell \tag{Tonelli} \\
        &= \int_0^\infty w_y(\ell)^\alpha \, d\ell. 
    \end{align*}
    Together, we have shown the pointwise bound
    \begin{equation}
        \int_0^\infty w_y(\ell)^\alpha \, d\ell \leq \int_{\ZSpace} q_y(z)^{\alpha - 1} \, \,d\nu_y(z). \label{eq:pointwisebound} 
    \end{equation}
    Integration with respect to $P_Y$, taking logarithms, and dividing by $1-\alpha > 0$ (all of which preserve the inequality) give that
    \begin{align*}
        \RFI{\alpha}{X}{Y} &= \frac{1}{1-\alpha} \log \Exp_Y \left[\int_0^\infty w_Y(\ell)^\alpha \, d\ell  \right] \\
        &\leq \frac{1}{1-\alpha} \log \int_{\YSpace} \int_{\ZSpace} q_y(z)^{\alpha - 1} \nu_y(dz) \, dP_Y(y) \\
        &= \frac{1}{1-\alpha} \log \Exp_Z \left[ \sum_{y} q_y(Z)^\alpha \right] \\
        &= \frac{\alpha}{1-\alpha} \log \left[ \left( \Exp_Z \left[ \sum_{y} q_y(Z)^\alpha \right] \right)^{\frac{1}{\alpha}} \right] \\
        &\leq \frac{\alpha}{1-\alpha} \log  \Exp_Z \left(
          \left[\sum_{y} q_y(Z)^\alpha \right]^{\frac{1}{\alpha}}
          \right) \\
        &= H^{\mathcal{A}}_\alpha(Y \mid Z), 
    \end{align*}
    where the inequality follows since  $x \mapsto
    x^{\frac{1}{\alpha}}$  is convex if $\alpha \in (0, 1)$.     
    We now turn to showing the upper bound in~\eqref{eq:renyi_pfr_theorem}. Specifically, for $Z = \qty{(Y_i, T_i)}_{i \ge 1}$ the Poisson process independent of $X$ with intensity measure $Q_Y \otimes \lambda$, where $P_{Y\mid X=x}\ll Q_Y$ for $P_X$-a.e.~$x$, we will show that the output $Y=Y_K$ of the Poisson functional representation in Theorem~\ref{thm:rtc_for_pfr} satisfies 
    \begin{equation}
        H_\alpha^{\mathcal{A}}(Y \mid Z) \le \RFI{\alpha}{X}{Y} + \frac{1 - \alpha}{\alpha} + \frac{\log (1 / \alpha)}{1 - \alpha}. \label{eq:renyi_bound_to_show}
    \end{equation}
    This gives the upper bound since $R^*_\alpha(X\to Y)\le H_\alpha^{\mathcal A}(Y\mid Z)$. \par
    Multiplying both sides of~\eqref{eq:renyi_bound_to_show} by $1 - \alpha > 0$ and exponentiating, this inequality is equivalent to
    \begin{equation}
        \left(\Exp_Z \norm{P_{Y \mid Z}}_{\alpha}\right)^\alpha \le
        \frac{2^{(1 - \alpha)^2 / \alpha}}{\alpha} 2^{(1 - \alpha)
          \RFI{\alpha}{X}{Y}},  \label{eq:eq_to_show} 
    \end{equation}
    Here $\norm{P}_\alpha=(\sum_u P(u)^\alpha)^{1/\alpha}$ for a purely atomic  $P$, with the sum over its atoms. By subadditivity of $x\mapsto x^\alpha$, for $P_Z$-a.e.~$z$ we have
    \begin{equation}
        \norm{P_{Y \mid Z = z}}_{\alpha} \le \norm{P_{K \mid Z = z}}_{\alpha}. \label{eq:correct_norm}
    \end{equation}
    As we do not know that $\Exp_Z \norm{P_{K \mid Z}}_\alpha$ is finite a priori, fix $L > 0$ and write $\norm{P_{K \mid Z}}_\alpha \land L = \min \qty{\norm{P_{K \mid Z}}_\alpha, L}$. Then, 
    \begin{align}
        \MoveEqLeft \Exp_Z \left[ \norm{P_{K \mid Z}}_\alpha \land L \right] \nonumber \\
        &\le \Exp_{Z} \left[ \norm{P_{K \mid Z}}_\alpha^\alpha \left( \norm{P_{K \mid Z}}_\alpha \land L \right)^{1-\alpha} \right] \nonumber \\
        &= \Exp_{Z} \left[ \sum_{k} P_{K \mid Z}(k \mid Z)^\alpha \left( \norm{P_{K \mid Z}}_\alpha \land L \right)^{1-\alpha}\right]. \label{eq:mecke_sum}
    \end{align}
    For common randomness $Z = z$ and a point $(y, t)$, define
    \begin{equation}
        S_{y, t}(z) = \qty{ x : \frac{t}{r_Q(y \mid x)} \le \min_{(y', t') \in z} \frac{t'}{r_Q(y' \mid x)}}
    \end{equation}
    to be the inputs for which $(y,t)$ would be the returned point for the common randomness $z\cup\{(y,t)\}$, up to ties, which have zero measure in the integral below.
    We can apply Mecke's identity to the sum in~\eqref{eq:mecke_sum} to yield that
    \begin{align}
       &\Exp_{Z} \left[ \sum_{k} P_{K \mid Z}(k \mid Z)^\alpha \left( \norm{P_{K \mid Z}}_\alpha \land L \right)^{1-\alpha}\right] \nonumber \\*
       &= \!\int_{\YSpace} \!\int_0^\infty \!\!\Exp_{Z}\!\! \left[
         P_X(S_{y, t}(Z))^\alpha \!\left( \norm{P_{K \mid Z}(\cdot\mid Z\cup\{(y,
         t)\})}_\alpha \!\!\land\! L \right)^{1\!-\!\alpha}\!\right]
         \!dt\,  dQ_Y(y).
    \end{align}
    For the extended common randomness $z\cup\{(y,t)\}$, give the inserted point index $0$ and retain the original indices of the other points. This relabeling preserves the norm. Then,
    \begin{align}
        \MoveEqLeft \norm{P_{K \mid Z = z \cup \{(y, t)\}}}_\alpha^\alpha \nonumber \\
        &= P_X(S_{y, t}(z))^\alpha + \sum_{k} \Prob(K = k, X \notin S_{y, t}(z) \mid Z = z)^\alpha  \nonumber \\
        &\le \sum_{k} \big( \Prob(K = k, X \in S_{y, t}(z) \mid Z = z)^\alpha \nonumber \\
        &\quad\quad\quad + \Prob(K = k, X \notin S_{y, t}(z) \mid Z = z)^\alpha \big) \label{eq:subadditivity} \\
        &\le 2^{1-\alpha} \sum_k \Prob(K = k \mid Z = z)^\alpha  \label{eq:other_inequality} \\
        &= 2^{1-\alpha} \norm{P_{K \mid Z=z}}_\alpha^\alpha. \nonumber
    \end{align}
    Above,~\eqref{eq:subadditivity} holds by the subadditivity of $x \mapsto x^\alpha$ and~\eqref{eq:other_inequality} holds by $a^\alpha + b^\alpha \le 2^{1-\alpha} (a + b)^\alpha$ for $a, b \ge 0$. Therefore, 
    \begin{align}
        \MoveEqLeft \Exp_{Z} \left[ \norm{P_{K \mid Z}}_\alpha \land L \right] \nonumber \\
        &\le \!\int_{\YSpace} \!\int_0^\infty \!\!\!\Exp_{Z}\!\!
          \left[ P_X(S_{y, t}(Z))^\alpha \!\left( 2^{\frac{1 -
          \alpha}{\alpha}} \norm{P_{K \mid Z}}_\alpha \!\!\land\! L
          \!\right)^{1\!-\!\alpha}\!\right] \!dt \, dQ_Y(y) \nonumber \\
        &\le 2^{\frac{(1 - \alpha)^2}{\alpha}} \!\Exp_Z \!\!\left[
          \int_{\YSpace}\! \int_0^\infty \!\!\!\!\!P_X(S_{y,
          t}(Z))^\alpha \, dt\,  dQ_Y(y) \!\left( \norm{P_{K \mid Z}}_\alpha \!\!\land\! L\! \right)^{1\!-\!\alpha} \!\right], \nonumber
    \end{align}
    where the last inequality uses $(ca)\land L\le c(a\land L)$ for $c\ge1$, $a\ge0$, and Tonelli's theorem. We next apply H\"older's and Minkowski's inequalities to obtain
\begin{align}
\MoveEqLeft \Exp_Z \!\left[ \int_{\YSpace}\! \int_0^\infty \!\!\!P_X(S_{y, t}(Z))^\alpha \,dt\, dQ_Y(y) \!\left( \norm{P_{K \mid Z}}_\alpha \!\land\! L \right)^{1-\alpha} \right] \nonumber \\
&\le \Exp_Z \left[ \left(\int_{\YSpace} \int_0^\infty P_X(S_{y,
                                                                                                                                                                                       t}(Z))^\alpha \, dt \, dQ_Y(y) \right)^{\frac{1}{\alpha}} \right]^{\alpha} \nonumber \\
&\quad \quad \times \left( \Exp_Z \left[ \norm{P_{K \mid Z}}_\alpha \land L \right] \right)^{1-\alpha} \nonumber \\
&\le \left(\int_{\YSpace} \int_0^\infty \Exp_Z \left[P_X(S_{y,
                                                                                                                      t}(Z))\right]^\alpha \, dt \, dQ_Y(y) \right) \nonumber \\
&\quad\quad \times \left( \Exp_Z \left[ \norm{P_{K \mid Z}}_\alpha \land L \right] \right)^{1-\alpha}. \nonumber 
\end{align}
    Above, the first inequality is Hölder's inequality with conjugate exponents $1/\alpha$ and $1/(1-\alpha)$, and the second inequality is Minkowski's integral inequality with exponent $1 / \alpha$. \par
    Following the discussion in Section~\ref{sec:upper_pfr}, we have that
    \begin{equation*}
        \Exp_{Z} \left[ P_X(S_{y, t}(Z)) \right] = a_y^Q(t) = \int_{\XSpace} \exp\left( - \frac{t}{r_Q(y \mid x)} \right) \, dP_X(x).
    \end{equation*}
    Therefore, dividing both sides of our original inequality by $\left( \Exp_Z \left[ \norm{P_{K \mid Z}}_\alpha \land L \right] \right)^{1-\alpha}$, which is finite and strictly positive, and letting $L\uparrow\infty$ by monotone convergence, we obtain from~\eqref{eq:correct_norm} 
    \begin{equation}
        \left(\Exp_Z \norm{P_{Y \mid Z}}_{\alpha}\right)^\alpha \le
        2^{\frac{(1-\alpha)^2}{\alpha}} \int_{\YSpace} \int_0^\infty
        a_y^Q(t)^\alpha \, dt \, dQ_Y(y). \label{eq:almost_there}
    \end{equation}
    Finally, we claim that, for $Q_Y$-a.e.~$y$,
    \begin{equation}
        \int_0^\infty a_y^Q(t)^\alpha \, dt \le \frac{1}{\alpha} \int_0^\infty w_y^Q(\ell)^\alpha \, d\ell. \label{eq:final_width_inequality}
    \end{equation}
    To show~\eqref{eq:final_width_inequality}, fix $y$, $t > 0$, and note that, for every $x$,
    \begin{equation}
        \exp \left(-\frac{t}{r_Q(y \mid x)}\right) = \left( \int_0^\infty \Ind{s r_Q(y \mid x) \ge t}^\alpha \alpha e^{-\alpha s} \, ds \right)^{1/\alpha}.
    \end{equation}
    Thus, 
    \begin{align}
        \MoveEqLeft  a_y^Q(t) \nonumber \\
        &\le \left( \int_0^\infty \left(\int_{\XSpace} \Ind{s r_Q(y \mid x) \ge t}\, dP_X(x) \right)^{\alpha} \alpha e^{-\alpha s} \, ds \right)^{1/\alpha} \nonumber \\
        &= \left(\int_0^\infty \alpha e^{-\alpha s} w_y^Q(t / s)^\alpha \, ds \right)^{1/\alpha}. \nonumber
    \end{align}
    Above, the inequality follows from the reverse Minkowski integral inequality for nonnegative functions with $0<\alpha<1$ (equivalently, apply Minkowski in $L^{1/\alpha}(P_X)$ to the integral in $s$ before taking the $1/\alpha$ power). Raising both sides to the power of $\alpha$ and integrating over $t$ yields
    \begin{align*}
        \int_0^\infty a_y^Q(t)^\alpha \, dt &\le \int_0^\infty \alpha s e^{-\alpha s} \, ds \int_0^\infty w_y^Q(\ell)^\alpha \, d\ell \\
        &= \frac{1}{\alpha} \int_0^\infty w_y^Q(\ell)^\alpha \, d\ell.
    \end{align*} Substituting~\eqref{eq:final_width_inequality} into~\eqref{eq:almost_there} yields
    \begin{equation}
        \left( \Exp_Z \norm{P_{Y \mid Z}}_\alpha \right)^\alpha \le \frac{2^{(1 - \alpha)^2 / \alpha}}{\alpha}  \left( \int_{\YSpace} \int_0^\infty w_y^Q(\ell)^\alpha \, d\ell \, dQ_Y(y) \right),
    \end{equation}
    from which the result follows by the change of variables below. Where $\RND{P_Y}{Q_Y}(y)=0$, $w_y^Q(\ell)=0$ for $\ell>0$, and the integrand is taken to be zero there. Thus,
    \begin{align}
        \MoveEqLeft \int_{\YSpace} \int_0^\infty w_y^Q(\ell)^\alpha \, d\ell \, dQ_Y(y) \nonumber \\
        &= \int_{\YSpace} \int_0^\infty w_y \left( \frac{\ell}{\RND{P_{Y}}{Q_Y}(y)} \right)^\alpha \, d\ell \, dQ_Y(y) \nonumber \\
        &= \int_{\YSpace} \int_0^\infty w_y \left( h \right)^\alpha \RND{P_Y}{Q_Y}(y) \, dh \, dQ_Y(y)  \tag{sub. $h = \ell / \RND{P_Y}{Q_Y}(y)$} \\
        &= \int_{\YSpace} \int_0^\infty w_y \left( h \right)^\alpha \, dh \, dP_Y(y). \tag*{\qedhere}
    \end{align}
\end{proof}

\bibliographystyle{IEEEtran}
\bibliography{references}

\end{document}